\documentclass[a4paper,12pt]{article}
\pdfoutput=1
\usepackage{graphicx,rotating,hyperref,slashed,verbatim,amsmath,xcolor,amssymb,amsfonts,expdlist,colortbl,cite,youngtab}
\usepackage{multirow}
\usepackage{booktabs}    
\usepackage{calc}        

\hypersetup{colorlinks,bookmarksopen,bookmarksnumbered,
linkcolor=blus,pdfstartview=FitH,urlcolor=rossos,citecolor=verde}

\newcommand{\vialamerda}[1]{}
\newcommand{\BR}{\hbox{BR}}

\def\lsim{\mathrel{\rlap{\lower3pt\hbox{\hskip0pt$\sim$}}
   \raise1pt\hbox{$<$}}}         
\def\gsim{\mathrel{\rlap{\lower4pt\hbox{\hskip1pt$\sim$}}
   \raise1pt\hbox{$>$}}}         

 \newcommand{\sfootnote}[1]{}
\definecolor{bluc}{cmyk}{1,1,0,0.1}
\definecolor{rossoCP3}{cmyk}{0,.88,.77,.40}
\definecolor{rosso}{cmyk}{0,1,1,0.4}
\definecolor{giallo}{cmyk}{0,.33,1,0}
\definecolor{rossos}{cmyk}{0,1,1,0.55}
\definecolor{rossoc}{cmyk}{0,1,1,0.2}
\definecolor{verdes}{cmyk}{0.92,0,0.59,0.4}

\hypersetup{colorlinks, bookmarksopen, bookmarksnumbered,
citecolor=verdes, linkcolor=bluc, pdfstartview=FitH, urlcolor=rossos}

\renewcommand{\Re}{{\rm Re}\,}
\newcommand{\mio}[1]{}

\newcommand{\fig}[1]{~\ref{fig:#1}}

\definecolor{Gray}{gray}{0.95}
\newcommand{\bbox}[1]{\fcolorbox{gray}{Gray}{~$\displaystyle #1$~}}

\newcommand{\F}{{\cal F}}

\renewcommand{\S}{{\cal S}}

\newcommand{\sfrac}[2]{#1/#2}

\usepackage{multicol}
\usepackage{color}
\definecolor{rosso}{cmyk}{0,1,1,0.4}
\definecolor{rossos}{cmyk}{0,1,1,0.55}
\definecolor{rossoc}{cmyk}{0,1,1,0.2}
\definecolor{blu}{cmyk}{1,1,0,0.3}
\definecolor{blus}{cmyk}{1,1,0,0.6}
\definecolor{bluc}{cmyk}{1,1,0,0.1}
\definecolor{verde}{cmyk}{0.92,0,0.59,0.25}
\definecolor{verdec}{cmyk}{0.92,0,0.59,0.15}
\definecolor{verdes}{cmyk}{0.92,0,0.59,0.4}

\newcommand{\eq}[1]{~{\rm (\ref{eq:#1})}}

\newcommand{\MeV}{\,{\rm MeV}}
\newcommand{\GeV}{\,{\rm GeV}}
\newcommand{\TeV}{\,{\rm TeV}}

\newcommand{\diag}{\,{\rm diag}}

\def\circa#1{\,\raise.3ex\hbox{$#1$\kern-.75em\lower1ex\hbox{$\sim$}}\,}

\newcommand{\beq}{\begin{equation}}
\newcommand{\eeq}{\end{equation}}

\newcommand{\bea}{\begin{eqnarray}}
\newcommand{\eea}{\end{eqnarray}}
\newcommand{\be}{\begin{equation}}
\newcommand{\ee}{\end{equation}}
\font\tenrsfs=rsfs10 at 12pt
\font\sevenrsfs=rsfs7 at 10 pt
\font\fiversfs=rsfs5
\newfam\rsfsfam
\textfont\rsfsfam=\tenrsfs
\scriptfont\rsfsfam=\sevenrsfs
\scriptscriptfont\rsfsfam=\fiversfs
\def\mathscr#1{{\fam\rsfsfam\relax#1}}

\def\Lag{\mathscr{L}}

\def\circa#1{\,\raise.3ex\hbox{$#1$\kern-.75em\lower1ex\hbox{$\sim$}}\,}
\makeatletter

\def\hhref#1{\href{http://arxiv.org/abs/#1}{arXiv:#1}} 

\newcommand{\doi}[1]{\href{http://dx.doi.org/#1}{[doi]}}

\def\hhref#1{\href{http://arxiv.org/abs/#1}{arXiv:#1}}

\def\art{\@ifnextchar[{\eart}{\oart}}
\def\eart[#1]#2#3#4#5#6{{\rm #2}, {\em #3 \bf #4} {\rm (#6) #5} ({\em #1})}

\def\article{\@ifnextchar[{\earticle}{\oarticle}}
\def\oarticle#1#2#3#4#5#6{{\rm #1}, {\em ``#6''}, {\rm #2 #3 (#5) #4}}
\def\earticle[#1]#2#3#4#5#6#7{{\rm #2}, {\em ``#7''}, {\rm #3 #4 (#6) #5}  [\hhref{#1}]}
\def\hepart[#1]#2{{\rm #2, \em#1}}
\def\heparticle[#1]#2#3{#2, {\em ``#3''} [\hhref{#1}]}

\newcounter{alphaequation}[equation]
\def\thealphaequation{\theequation\hbox to
0.6em{\hfil\alph{alphaequation}\hfil}}
\def\eqnsystem#1{
\def\@eqnnum{{\rm (\thealphaequation)}}
\def\@@eqncr{\let\@tempa\relax \ifcase\@eqcnt \def\@tempa{& & &} \or
  \def\@tempa{& &}\or \def\@tempa{&}\fi\@tempa
  \if@eqnsw\@eqnnum\refstepcounter{alphaequation}\fi
\global\@eqnswtrue\global\@eqcnt=0\cr}
\refstepcounter{equation} \let\@currentlabel\theequation \def\@tempb{#1}
\ifx\@tempb\empty\else\label{#1}\fi
\refstepcounter{alphaequation}
\let\@currentlabel\thealphaequation
\global\@eqnswtrue\global\@eqcnt=0 \tabskip\@centering\let\\=\@eqncr
$$\halign to \displaywidth\bgroup \@eqnsel\hskip\@centering
$\displaystyle\tabskip\z@{##}$&\global\@eqcnt\@ne
\hskip2\arraycolsep\hfil${##}$\hfil& \global\@eqcnt\tw@\hskip2\arraycolsep
$\displaystyle\tabskip\z@{##}$\hfil
\tabskip\@centering&\llap{##}\tabskip\z@\cr}
\def\endeqnsystem{\@@eqncr\egroup$$\global\@ignoretrue} \makeatother

\newcommand{\SU}{\,{\rm SU}}

\newcommand{\U}{\,{\rm U}}

\definecolor{fiorentina}{rgb}{.5,0,.5}

\allowdisplaybreaks

\begin{document}
\centerline{CERN-{TH}-2017-086,\hfill  IFUP-TH/2017\hfill CP3-Origins-2017-014}
 \vspace{1truecm}

\begin{center}
\boldmath

{\textbf{\LARGE\color{rossoCP3}
Flavour 
anomalies after the $R_{K^*}$ measurement }}

\centerline{\large\color{rossoCP3}\bf(updated including the Moriond 2019 data from LHCb and {\sc Belle}\footnote{The addendum at pages \pageref{6in}--\pageref{6out}
(section~6) is not present in the published version of this paper.})}

\unboldmath

\bigskip\bigskip

\large

{\bf Guido D'Amico$^a$, Marco Nardecchia$^a$, Paolo Panci$^a$, \\
 Francesco Sannino$^{a,b}$,  Alessandro Strumia$^{a,c}$, \\Riccardo Torre$^d$, Alfredo Urbano$^a$} \\[8mm]
 {\it $^a$ Theoretical Physics Department, CERN, Geneva, Switzerland}\\
{\it $^b$ CP$^3$-Origins and Danish IAS, University of Southern Denmark, Denmark}\\[1mm]
{\it $^c$ Dipartimento di Fisica dell'Universit{\`a} di Pisa and INFN, Italy}\\[1mm]
{\it $^d$ Theoretical Particle Physics Laboratory, Institute of Physics, EPFL, Lausanne, Switzerland}\\[1mm]


\bigskip\bigskip

\thispagestyle{empty}\large
{\bf\color{blus} Abstract}
\begin{quote}
The LHCb measurement of the $\mu/e$ ratio $R_{K^*}$ indicates a deficit with respect to the Standard Model prediction, supporting earlier hints of lepton universality violation observed in the $R_K$ ratio.
We show that the  $R_K$ and $R_{K^*}$ ratios alone constrain the chiralities of the states  contributing to these anomalies, and we find deviations from the Standard Model at the $4\sigma$ level.
This conclusion is further corroborated by  hints from the theoretically challenging  $b\to s\mu^+\mu^-$ distributions.
Theoretical interpretations  in terms of $Z'$, lepto-quarks, loop mediators, and composite dynamics  are discussed.
We highlight their distinctive features in terms of the chirality and flavour structures relevant to the observed anomalies.
\end{quote}
\thispagestyle{empty}
\end{center}

\setcounter{page}{1}
\setcounter{footnote}{0}


\newpage
\tableofcontents

\section{Introduction}
 The LHCb~\cite{RKstar} collaboration presented their results on the measurement of the ratio
\beq\label{eq:RK*def}
R_{K^*}=\frac{\hbox{BR}(B \to K^* \mu^+ \mu^-) }{ \hbox{BR}(B \to K^* e^+ e^-) } \, .
\eeq
The aim of this measurement is to test the universality of the gauge interactions in the lepton sector.
Taking the ratio of branching ratios strongly reduces the Standard Model (SM) theoretical uncertainties, as suggested for the first time in ref.~\cite{Hiller:2003js}.

The experimental result~\cite{RKstar} is reported in two bins of di-lepton invariant mass
\beq\label{eq:RK*LHCB}
R_{K^*}  = \left\{\begin{array}{ll}
  0.660^{+0.110}_{-0.070} \pm 0.024 & (2m_\mu)^2 < q^2 < 1.1 \GeV^2  \\[2mm]
  0.685^{+0.113}_{-0.069} \pm 0.047& 1.1  \GeV^2 < q^2 < 6   \GeV^2 \, .
\end{array}
\right.
\eeq
These values have to be compared with the SM predictions \cite{1605.07633}
\beq
R^{SM}_{K^*}  = \left\{\begin{array}{ll}
  0.906 \pm 0.028 & (2m_\mu)^2 < q^2 < 1.1 \GeV^2  \\[2mm]
  1.00 \pm 0.01 & 1.1  \GeV^2 < q^2 < 6   \GeV^2 \, .
\end{array}
\right.
\eeq

At face value, a couple of observables featuring a $\sim2.5\sigma$ deviation from the SM predictions  can be attributed to a mere statistical fluctuation.
The interest resides in the fact that such results might be part of a coherent picture involving New Physics~(NP)  in the $b \to s \mu^+ \mu^-$ transitions.
In fact, anomalous deviations were also observed in the following related measurements:
\begin{enumerate}
\item the $R_{K}$ ratio~\cite{1406.6482}
\beq \label{eq:RKexp}
R_K =\frac{{\rm BR} \left ( B^+ \to K^+ \mu ^+ \mu ^-\right )}{{\rm BR} \left ( B^+ \to K^+ e^+ e^-\right )}= 0.745\pm 0.09_{\rm stat} \pm0.036_{\rm syst} \, ;
\eeq
\item the branching ratios of the semi-leptonic decays $B \to K^{(*)} \mu^+ \mu^-$~\cite{1403.8044} and $B_s \to \phi \mu^+ \mu^-$~\cite{1506.08777};
\item the angular distributions of the decay rate of $B \to K^* \mu^+ \mu^-$.
In particular, the so-called $P'_5$ observable (defined for the first time in \cite{1207.2753}) shows the most significant discrepancy~\cite{1308.1707,1403.8044,1512.04442}.
\end{enumerate}

The coherence of this pattern of deviations has been pointed out already after the measurement of $R_K$ with a subset of observables in  \cite{1408.1627, 1408.4097} and in a full global analysis in \cite{1411.3161,1510.04239}.

For the observables in points $2$ and $3$ the main source of uncertainty is theoretical.
It resides in the proper evaluation of the form factors and in the estimate of the non-factorizable hadronic corrections.
Recently, great theoretical effort went into the understanding of these aspects, see ref.s~\cite{1006.4945,1101.5118,1207.2753,1211.0234,1212.2263,1406.0566,1407.8526,1412.3183,1512.07157,1701.08672,1702.02234}  for an incomplete list of references.

Given their reduced sensitivity to theoretical uncertainties in the SM, the $R_K$ and $R_{K^*}$  observables offer a neat way to establish potential violation of lepton flavour universality. Future data will be able to further reduce the statistical uncertainty on these quantities.
In addition, measurements of other ratios $R_H$ analogous to $R_K$, with $H=X_s, \phi, K_0 (1430),f_0$ will constitute relevant independent tests \cite{Hiller:2003js,1411.4773}.

The paper is structured as follows.
In section~\ref{obs} we discuss the relevant observables and how they are affected by additional effective operators.
We perform a global fit  in section~\ref{fit}.
We show that, even restricting the analysis to the theoretically clean $R_K$, $R_{K^*}$  ratios, the overall deviation from the SM starts to be significant, at the $4 \sigma$ level, and to point towards some model building directions.
Such results prompt us to investigate, in section~\ref{th}, a few theoretical interpretations.
We discuss models including $Z'$,  lepto-quark exchanges, new states affecting the observables via quantum corrections, and models of composite Higgs.

\section{Effective operators and observables}\label{obs}
Upon integrating out heavy degrees of freedom the relevant processes can be described, near the Fermi scale, in terms of the effective Lagrangian
\beq\label{eq:Leff}
\mathscr{L}_{\rm eff}=
\sum_{\ell, X, Y} c_{b_X \ell_Y}  {\cal O}_{b_X \ell_Y}\eeq
where the sum runs over leptons $\ell =   \{e,\mu , \tau\}$ and over their chiralities $X,Y = \{L,R\}$.
New physics is more conveniently explored in  the chiral basis
\beq {\cal O}_{b_X \ell_Y} = (\bar s \gamma_\mu P_X b)(\bar \ell \gamma_\mu P_Y \ell).\eeq
These vector operators can be promoted to $\SU(2)_L$-invariant operators, unlike scalar or tensor operators~\cite{1407.7044}.
In SM computations one uses the equivalent formulation
\beq \mathscr{H}_{\rm eff}
=  - V_{tb} V_{ts}^*  \frac{\alpha_{\rm em}}{4\pi v^2}
\sum_{\ell, X, Y} C_{b_X \ell_Y}  {\cal O}_{b_X \ell_Y}\, + \mathrm{h.c.} \, , \eeq
defining dimensionless coefficients $C_I$ as
\beq c_I = V_{tb} V_{ts}^*  \frac{\alpha_{\rm em}}{4\pi v^2} C_I   = - \frac{C_I}{(36\TeV)^2} \, ,
\eeq
where
$V_{ts}= - 0.040\pm 0.001$ has a negligible imaginary part,
$v = 174\GeV$ is the Higgs vacuum expectation value, usually written as $1/v^2 = 4 G_{\rm F}/\sqrt{2}$.
The SM itself contributes as $C^{\rm SM}_{b_L \ell_L} = 8.64$ and $C^{\rm SM}_{b_L \ell_R} = -0.18$, accidentally implying $|C^{\rm SM}_{b_L \ell_R}| \ll |C^{\rm SM}_{b_L \ell_L}|$.

This observation suggests  to use the chiral basis, related to the conventional one (see e.g. ref.~\cite{1411.3161}) by $C_9 = C_{b_L \mu_{L+R}}/2$, $C_{10}=-C_{b_L \mu_{L-R}}/2$, $C'_9 = C_{b_R \mu_{L+R}}/2$, $C'_{10}=-C_{b_R \mu_{L-R}}/2$, with the approximate relation $C_9^{\rm SM} \approx - C_{10}^{\rm SM}$ holding in the SM.
To make the notation more compact, we define
$C_{b_{L\pm R} \ell_{Y}} \equiv C_{b_L \ell_Y} \pm C_{b_R\ell_Y}$ and
$C_{b_{L+R} \ell_{L\pm R}} \equiv C_{b_L \ell_L}+ C_{b_R \ell_L} \pm C_{b_L \ell_R} \pm C_{b_R \ell_R}$, and $C_{b_X (\mu-e)_Y} \equiv C_{b_{X} \mu_{Y}}- C_{b_{X} e_{Y}}$.

\medskip

We now summarize the theoretically clean observables\footnote{By theoretically clean observables we mean those ones predicted in the SM with an error up to few percent.}, presenting both the full expressions and the ones in chiral-linear approximation.
The latter is defined by neglecting $| C^{\rm SM}_{b_L \ell_R}| \ll | C^{\rm SM}_{b_L \ell_L}  | $ and
expanding each coefficient $C_I$ at first order in the beyond-the-standard-model (BSM) contribution, $C_I = C^{\rm SM}_I + C^{\rm BSM}_I$.

\subsection{$R_K$ revisited}
The experimental analysis is made by binning the observable in the squared invariant mass of the lepton system $q^2\equiv (P_{\ell^-} + P_{\ell^+})^2$.
Writing the explicit $q^2$-dependence, we have
\begin{equation}\label{eq:RKq2}
R_K[q_{\rm min}^2,q_{\rm max}^2]\equiv \frac{\int_{q_{\rm min}^2}^{q_{\rm max}^2}dq^2 d\Gamma(B^+ \to K^+ \mu^+ \mu^-)/dq^2}{\int_{q_{\rm min}^2}^{q_{\rm max}^2}dq^2 d\Gamma(B^+ \to K^+ e^+ e^-)/dq^2} \, .
\end{equation}
The experimental value cited in eq.~(\ref{eq:RKexp}) refers to $R_K \equiv R_K[1~{\rm GeV}^2,6~{\rm GeV}^2]$.
To simplify the notation, however, in the following we will omit the units in brackets.
Neglecting SM contributions from the electromagnetic dipole operator, justified by the cut $q_{\rm min}^2 = 1$ GeV$^2$,
and non-factorizable contributions from the weak effective Hamiltonian,\footnote{In the limit of vanishing lepton masses the decay
rate in eq.~(\ref{eq:RKq2}) takes the form~\cite{1411.3161}
\begin{equation}\label{eq:RKfull}
\frac{ d\Gamma(B^+ \to K^+ \mu^+ \mu^-)}{dq^2} = \frac{G_{\rm F}^2\alpha_{\rm em}^2|V_{tb}V_{ts}^*|^2}{2^{10}\pi^5 M_B^3}\lambda^{3/2}(M_B^2, M_K^2, q^2)
\left(
|F_V|^2 + |F_A|^2
\right)~,
\end{equation}
where $G_{\rm F}$ is the Fermi constant, $\lambda(a,b,c) \equiv a^2 + b^2 + c^2 -2(ab+bc+ac)$, $M_B \approx 5.279$ GeV, $M_K \approx  0.494$ GeV,
$|V_{tb}V_{ts}^*| \approx 40.58\times 10^{-3}$. Introducing the QCD form factors $f_{+,T}(q^2)$ we have
\begin{eqnarray}
F_A(q^2) &=& \left(C_{10} + C_{10}^{\prime}\right)f_+(q^2)~,\\
F_V(q^2) &=& (C_9 + C_9^{\prime})f_+(q^2) + \underbrace{\frac{2m_b}{M_B + M_K}\left(C_7 +
C_7^{\prime}\right)f_T(q^2)}_{\rm SM\,electromagnetic\,dipole\,contribution} + \underbrace{h_K(q^2)}_{\rm non-factorizable\,term}~.
\end{eqnarray}
Notice that for simplicity we wrote the Wilson coefficient $C_9$ omitting higher-order $\alpha_s$-corrections~\cite{1703.10005}.
Neglecting SM electromagnetic dipole contributions (encoded in the coefficients $C_7^{(\prime)}$),
and non-factorizable corrections, eq.~(\ref{eq:RKTheory}) follows from Eqs~(\ref{eq:RKq2},\ref{eq:RKfull})
by rotating the coefficients $C_{9,10}^{(\prime)}$ on to the chiral basis.
}
the theoretical prediction for  $R_K$ is
\beq \label{eq:RKTheory}
R_K 
= \frac{|C_{b_{L+R} \mu_{L-R}}|^2  + |C_{b_{L+R} \mu_{L+R}}|^2}{|C_{b_{L+R} e_{L-R}}|^2  + |C_{b_{L+R} e_{L+R}}|^2} \ .
\eeq
This is a clean observable, meaning that it is not affected by large theoretical uncertainties, and its SM prediction is $R_K=1$. QED corrections give a small departure from unity which, however, does not exceed few percents~\cite{1605.07633}.
However, it has to be noted that new physics which affects differently  $\mu$ and  $e$ can induce theoretical errors, bringing back the issue of hadronic uncertainties.

In the chiral-linear approximation, $R_K$ becomes
\beq\label{eq:RKChiral}
 \bbox{R_K \simeq 1 +2  \frac{ \Re C^{\rm BSM}_{b_{L+R} (\mu-e)_{L}}}{C^{\rm SM}_{b_L \mu_L}}} \ ,
\eeq
indicating that the dominant effect stems from couplings to left-handed leptons.
Any chirality of quarks works,
as long as it is not orthogonal to $L+R$, namely unless quarks are axial.

It is important to notice that the approximation in eq.~(\ref{eq:RKChiral}), although capturing the relevant physics,
is not adequate for a careful phenomenological analysis.
The same remark remains valid for the simplified expression proposed in ref.~\cite{1411.4773},  expanded up to quadratic terms in new physics coefficients.
The reason is that the expansion is controlled by the parameter $C^{\rm BSM}_{b_X l_Y}/C^{\rm SM}_{b_X l_Y}$,
 a number that is not always smaller than $1$. This is particularly true in the presence of new physics in the electron sector in which --- as we shall discuss in detail ---
 large values of the Wilson coefficients are needed to explain the observed anomalies.
 For this reason, all the results presented in this paper  make use of the full expressions for both $R_K$~\cite{1411.3161} and, as we shall discuss next, $R_{K^*}$.

\subsection{Anatomy of $R_{K^*}$}\label{sec:AnatomyOfRKStar}\label{RK*an}
\begin{figure}[t]
\minipage{0.5\textwidth}
  \includegraphics[width=0.95\textwidth]{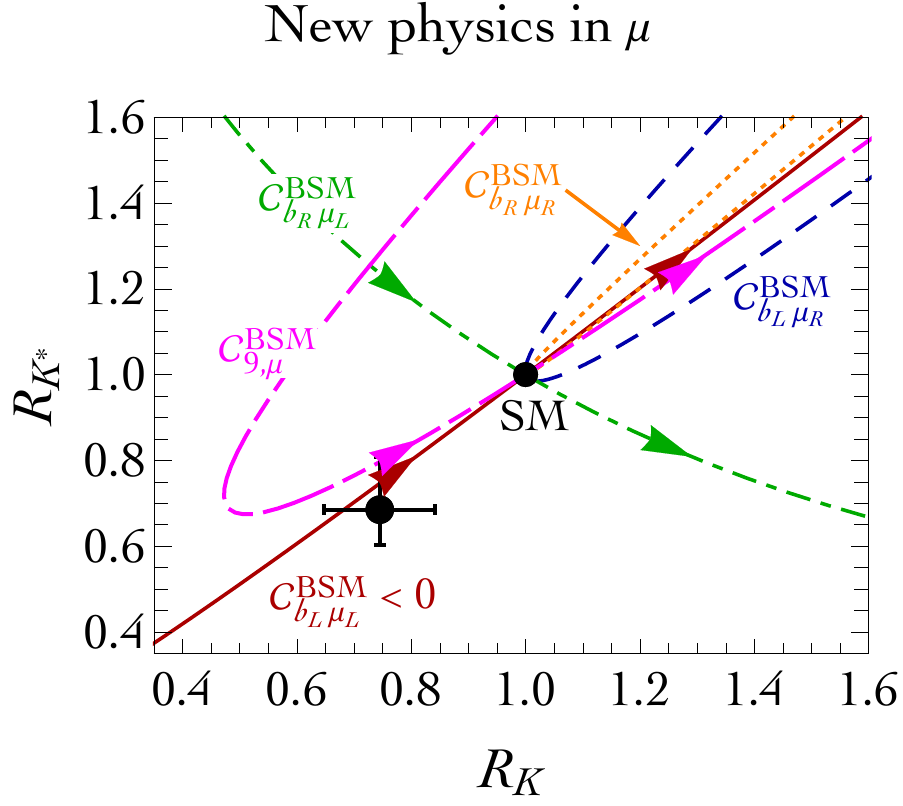}
\endminipage 
\minipage{0.5\textwidth}
  \includegraphics[width=0.95\textwidth]{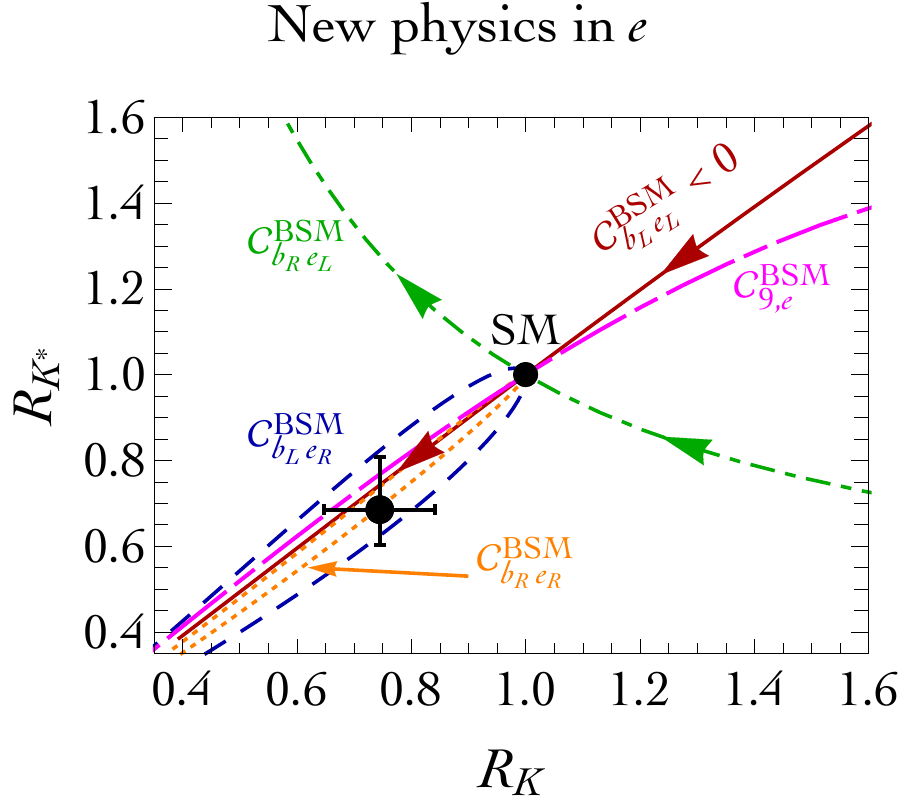}
\endminipage
\caption{\em Deviations from the SM value $R_K = R_{K^*} =1$ due to the various chiral operators possibly generated by new physics in the muon (left panel) and electron (right panel) sector.
Both ratios refer to the $[1.1,6] \GeV^{2}$ $q^2$-bin.
We assumed real coefficients, and the out-going (in-going) arrows show the effect of coefficients equal to $+1$ $(-1)$.
For the sake of clarity we only show the arrows for the coefficients involving left-handed muons and electrons (except for the two magenta arrows in the left-side plot, that refer to $C^{\rm BSM}_{9,\mu} = (C^{\rm BSM}_{b_L \mu_L} + C^{\rm BSM}_{b_L \mu_R})/2 = \pm 1$). The constraint from $B_s \to \mu \mu$ is not included in this plot.
\label{fig:RKRKStar}}
\end{figure}

Given that the $K^*$ has spin 1 and mass $M_{K^*} = 892\MeV$, the theoretical prediction for the $R_{K^*}$ ratio given in eq.\eq{RK*def} is
\beq
R_{K^*}
= \frac{(1-p) (|C_{b_{L+R} \mu_{L-R}}|^2  + |C_{b_{L+R} \mu_{L+R}}|^2 ) + p \left( |C_{b_{L-R} \mu_{L-R}}|^2  + |C_{b_{L-R} \mu_{L+R}}|^2 \right) }
{(1-p)( |C_{b_{L+R} e_{L-R}}|^2  + |C_{b_{L+R} e_{L+R}}|^2 ) + p \left( |C_{b_{L-R}e_{L-R}}|^2  + |C_{b_{L-R}e_{L+R}}|^2 \right)}
\eeq
 where $p\approx 0.86$ is the ``polarization fraction"~\cite{0805.2525,1308.4379,1411.4773}, that is defined as
 \begin{equation}
 p =\frac{g_0 +g_{\parallel}}{g_0 +g_{\parallel}+g_{\perp}} \, .
 \end{equation}
 The $g_i$ are the contributions to the decay rate (integrated over the the intermediate bin) of the different helicities of the $K^*$. The index $i$ distinguishes the various helicities: longitudinal ($i=0$), parallel ($i=\parallel$) and perpendicular ($i=\perp$).
In the  chiral-linear  limit the expression for  $R_{K^*}$ simplifies to
\beq \label{eq:RKStarChiral}
 \bbox{R_{K^*} \simeq R_K - 4p \frac{\Re C^{\rm BSM}_{b_R(\mu-e)_L}}{C^{\rm SM}_{b_L \mu_L}} }  \ ,
\eeq
where $4p/{C^{\rm SM}_{b_L \mu_L}}\approx 0.40$.
The formula above clearly shows that, in this approximation, a deviation of $R_{K^*}$ from $R_K$ signals that $b_R$ is involved at the effective operator level with the dominant effect still due to left-handed leptons.
As already discussed before, eq.~(\ref{eq:RKStarChiral}) is not suitable for a detailed phenomenological study, and we implement in our numerical code the full expression for $R_{K^*}$~\cite{0811.1214}.
In the left panel of figure~\ref{fig:RKRKStar},  we present the different predictions in the $(R_K, R_{K^*})$ plane due to turning on the various operators assumed to be generated via  new physics in the muon sector.
A reduction of the same order
in both $R_K$ and $R_{K^*}$ is possible in the presence of the left-handed operator
$C^{\rm BSM}_{b_L\mu_L}$ (red solid line).  In order to illustrate the size of the required correction, the arrows correspond to $C^{\rm BSM}_{b_L\mu_L} = \pm 1$ (see caption for details).
Conversely, as previously mentioned, a deviation of $R_{K^*}$ from $R_K$ signals the presence of $C^{\rm BSM}_{b_R\mu_L}$ (green dot-dashed line).
Finally, notice that the reduced value of $R_K$ measured in eq.~(\ref{eq:RKexp}) cannot be explained by $C^{\rm BSM}_{b_R\mu_R}$ and $C^{\rm BSM}_{b_L\mu_R}$.
The information summarized in this plot is of particular significance since it shows at a glance, and before an actual fit to the data, the new physics patterns implied by the combined measurement of $R_K$ and  $R_{K^*}$.

Before proceeding, another important comment is in order. In the left panel of figure~\ref{fig:RKRKStar}, we also show in magenta the direction
described by non-zero values of the coefficient $C^{\rm BSM}_{9,\mu} = (C^{\rm BSM}_{b_L \mu_L} + C^{\rm BSM}_{b_L \mu_R})/2$.
The latter refers to the effective operator $\mathcal{O}_{9}^{\mu} = (\bar{s}\gamma_{\mu}P_L b)( \bar{\mu}\gamma^{\mu}\mu)$, and implies a vector coupling for the muon.
The plot suggests that negative values $C^{\rm BSM}_{9,\mu}\approx -1$ may also provide a good fit of the observed data.
However, it is also interesting to notice that in the non-clean observables, the hadronic effects might mimic a short distance BSM contribution in $C^{\rm BSM}_{9,\mu}$.
From the plot in our figure \ref{fig:RKRKStar}, it is clear that with more data a combined analysis of $R_K$ and $R_{K^*}$ might start to discriminate between $C^{\rm BSM}_{9,\mu}$ and  $C^{\rm BSM}_{b_L \mu_L}$ using \textit{only} clean observables.
However, with the present data, there is only a mild preference for $C^{\rm BSM}_{b_L \mu_L}$, according to the 1-parameter fits of section \ref{fit:clean} using only clean observables.

It is also instructive to summarise  in the right panel of figure~\ref{fig:RKRKStar}  the case in which new physics directly affects the electron sector.  The result is a mirror-like image of the  muon case  since the coefficients $C_{b_X e_Y}$ enter, both at the linear and quadratic level, with an opposite sign when compared to their analogue $C_{b_X \mu_Y}$.
In the chiral-linear limit the only operator that can bring the values of
$R_K$ and  $R_{K^*}$ close to the experimental data is $C_{b_L e_L} > 0$. As before, a deviation from $R_K$ in  $R_{K^*}$ can
be produced by a non-zero value of $C^{\rm BSM}_{b_R e_L}$.
Notice that, beyond the chiral-linear limit,  also $C^{\rm BSM}_{b_{L,R} e_R}$  points towards the observed experimental data but
they require larger numerical values.

\begin{figure}[t]
\minipage{0.5\textwidth}
  \includegraphics[width=0.93\textwidth]{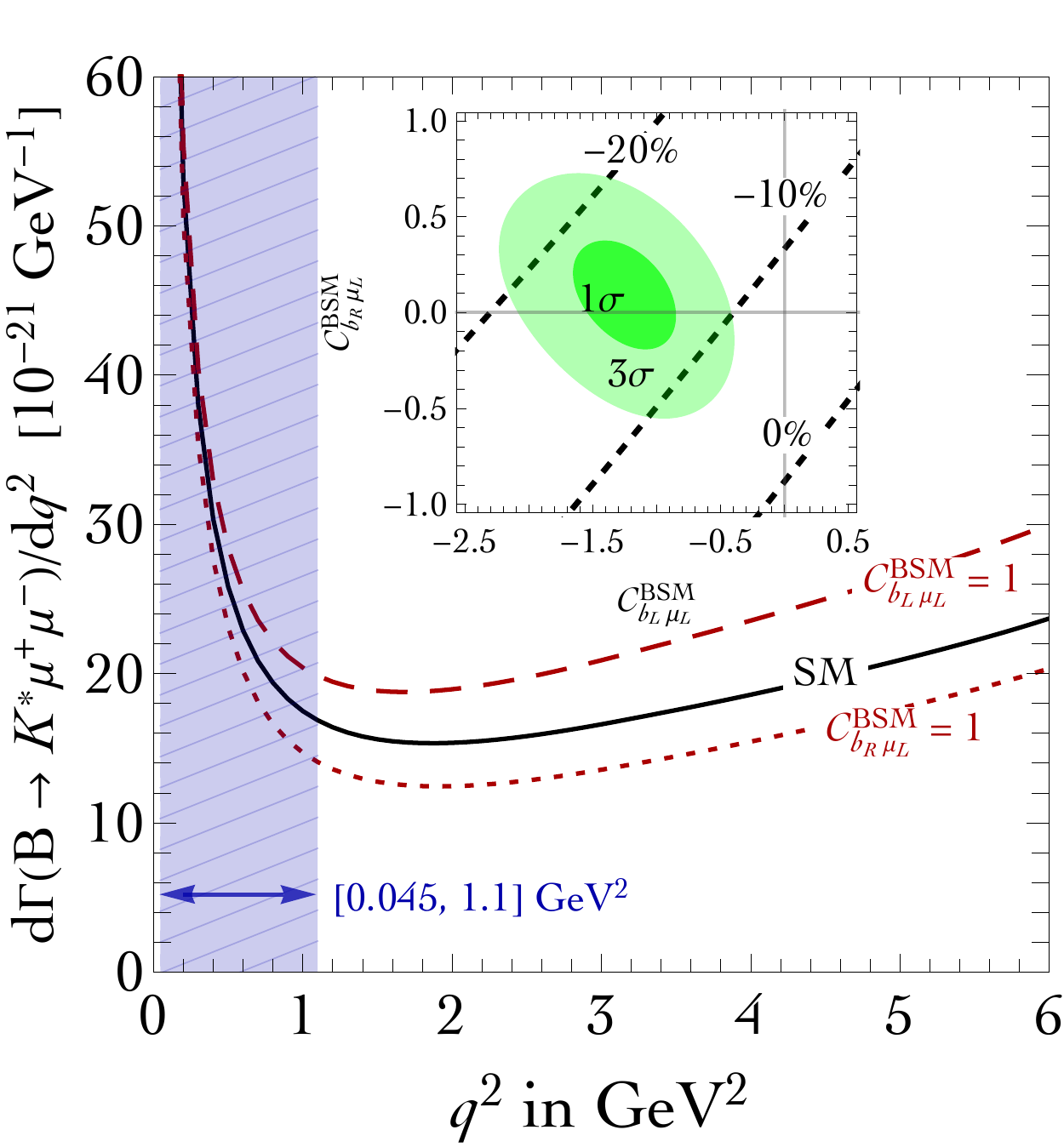}
\endminipage 
\minipage{0.5\textwidth}
  \includegraphics[width=0.96\textwidth]{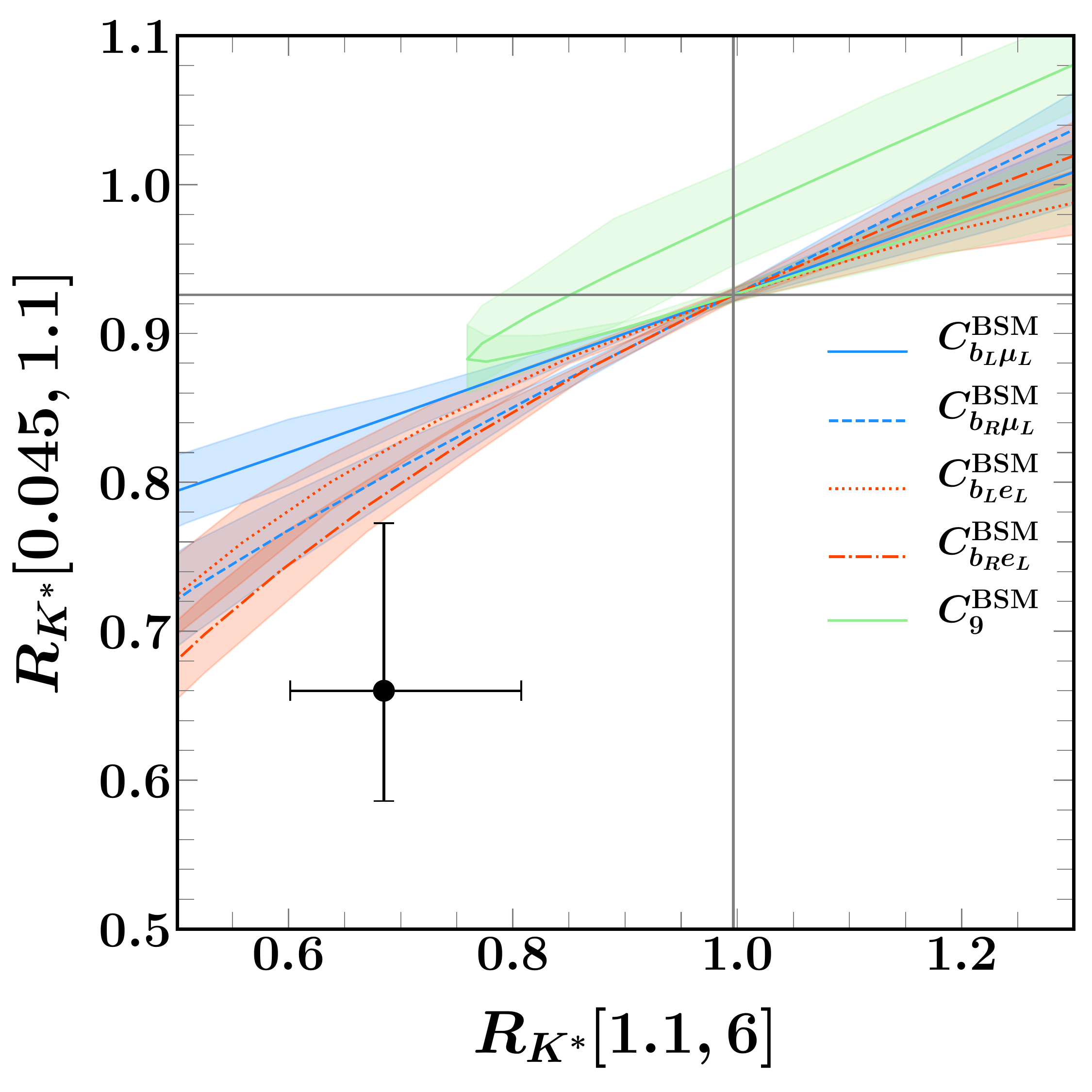}
\endminipage
\caption{\label{fig:RKAnatomy}\em
{\bf Left:} $R_{K^*}$ as function of $q^2$, the invariant mass of the $\ell^+\ell^-$ pair,
for the SM and for two specific values of the new-physics coefficients.
The inset shows iso-contours of deviation from $R_K^*=1$ in the $[0.045,1.1] \GeV^2$ bin as a function of new-physics coefficients, compared to their experimentally favoured values.
{\bf Right}: correlation between $R_{K^*}$ measured in the $[1.1,6] \GeV^2$ bin (horizontal axis) and $[0.045,1.1] \GeV^2$ bin (vertical axis) of $q^2$: a sizeable new physics effect can be present in the low-energy bin.
The numerical values of $q^2$ are given in $\GeV^2$.}
\end{figure}

\medskip

A closer look to $R_{K^*}$ reveals additional observable consequences related to the presence of BSM corrections.
$R_{K^*}$, in a given range of $q^2$, is defined in analogy with eq.~(\ref{eq:RKq2}):
\begin{equation}\label{eq:RKq}
R_{K^*}[q_{\rm min}^2, q_{\rm max}^2] \equiv \frac{\int_{q_{\rm min}^2}^{q^2_{\rm max}}dq^2\, \sfrac{d\Gamma(B \to K^* \mu^+ \mu^-)}{dq^2}}
{\int_{q_{\rm min}^2}^{q^2_{\rm max}} dq^2\, \sfrac{d\Gamma (B \to K^* e^+ e^-)}{dq^2}}~,
\end{equation}
where the differential decay width $d\Gamma(B \to K^* \mu^+ \mu^-)/dq^2$ actually describes the four-body process
$B \to K^*(\to K\pi) \mu^+ \mu^-$, and takes the compact form
\begin{equation}\label{eq:FullWidth}
\frac{d\Gamma \left(B \to K^* \mu^+ \mu^-\right)}{dq^2} = \frac{3}{4}\left(
2\mathcal{I}_{1}^{s} + \mathcal{I}_2^c
\right) - \frac{1}{4}\left(
2\mathcal{I}_{2}^{s} + \mathcal{I}_2^c
\right)~.
\end{equation}
The angular coefficients $\mathcal{I}_{i=1,2}^{a=s,c}$ in eq.~(\ref{eq:FullWidth}) can be written in terms of the so-called transversity  amplitudes
describing the decay $B\to K^*V^*$  with the $B$ meson decaying  to an on-shell $K^*$ and a virtual photon or $Z$ boson which later decays into a lepton-antilepton pair.
We refer to ref.~\cite{0811.1214} for a comprehensive description of the computation.
In the left panel of figure~\ref{fig:RKAnatomy} we show the differential distribution $d\Gamma(B \to K^* \mu^+ \mu^-)/dq^2$
as a function of the dilepton invariant mass $q^2$. The solid black line represents the SM prediction,
and we show in dashed (dotted) red
the impact  of BSM corrections due to the presence of non-zero  $C^{\rm BSM}_{b_L\mu_L}$ ($C^{\rm BSM}_{b_R\mu_L}$) taken at the benchmark value of $1$.

We now focus on the low invariant-mass range $q^2 = [0.045, 1.1]$ GeV$^2$, shaded in blue with diagonal
mesh in the left panel of fig~\ref{fig:RKAnatomy}.
In this bin, the differential rate is dominated by the SM photon contribution.
It is instructive to give more quantitative comments.
In the inset plot in the left panel of fig~\ref{fig:RKAnatomy}, we show in the plane $(C^{\rm BSM}_{b_L\mu_L}, C^{\rm BSM}_{b_R\mu_L})$ the
relative deviation in $R_{K^*}[0.045,1.1]$ compared to its SM value $R^{\rm SM}_{K^*} \approx 0.9$, and we superimpose the 1- and 3-$\sigma$ confidence contours
allowed by the fit of experimental data (without including $R_{K^*}$).
This comparison shows that a $10\%$ reduction of $R_{K^*}$  in the mass-invariant bin $q^2 = [0.045, 1.1]$ GeV$^2$ is expected from the experimental data.
The SM prediction, $R^{\rm SM}_{K^*}[0.045,1.1] \approx 0.9$, departs from one because of QED effects which distinguish between $m_\mu$ and $m_e$.
The observed central value $R^{\rm SM}_{K^*}[0.045,1.1]=0.66$ can be again explained with possible effects of new physics.
The natural suspect is a new physics contribution to the dipole operator, but it can be shown that this cannot be very large because of bounds coming from the inclusive process $B \to X_s \gamma$, see for example ref.~\cite{1608.02556}.
We can instead correlate the effect in $R^{\rm SM}_{K^*}[0.045,1.1]$ with  $R^{\rm SM}_{K^*}[1.1,6]$.
The results are shown in the right panel of figure~\ref{fig:RKAnatomy}. Here we learn that the new physics hypotheses predict values larger than the one observed in the data.
However, since the experimental error is quite large, precise measurements are needed to settle this issue.

In conclusion, the picture emerging from a simple inspection of the relevant formulas for $R_K$ and $R_{K^*}$ is very neat, and can be summarized as follows:

\begin{itemize}

\item[$\circ$] New physics in the muon sector can easily explain  the observed deficits in $R_K$,$R_{K^*}$,
and we expect a  preference for negative values of the operator involving a left-handed current, $C^{\rm BSM}_{b_L \mu_L}$.
Sizeable deviations of $R_{K^*}$ from $R_K$ signal non-zero values for $C^{\rm BSM}_{b_R \mu_L}$.

\item[$\circ$] New physics in the electron sector represents a valid alternative, and positive values of $C^{\rm BSM}_{b_L e_L}$ are favoured.
Sizeable deviations of $R_{K^*}$ from $R_K$ signal non-zero values for $C^{\rm BSM}_{b_R e_L}$. However invoking NP only the electronic channels does not allow to explain other anomalies in the muon sector such as the angular observables.

\item[$\circ$] There exists an interesting correlation between $R_{K^*}$ in the $q^2$-bin $[1.1,6]$ GeV$^2$ and $[0.045, 1.1]$ GeV$^2$.
At present, all the new physics hypothesis invoked tend to predicts larger value of $R_{K*}$ in the low bin than the one preferred by the data.

\end{itemize}
In section~\ref{fit} we shall corroborate this  qualitative picture with quantitative fits.

\subsection{$B_s \to \mu^+ \mu^-$}
The rate is predicted as
\beq \bbox{\frac{\textrm{BR}(B_s \to \mu^+ \mu^-) }{\textrm{BR}(B_s \to \mu^+ \mu^-) _{\rm SM}} =
\bigg|\frac{C_{b_{L-R}\mu_{L-R}}}{C^{\rm SM}_{b_{L-R}\mu_{L-R}}}\bigg|^2} \ , \eeq
where
$
\textrm{BR}(B_s \to \mu^+ \mu^-)_{\textrm{SM}} = (3.65 \pm 0.23) \times 10^{-9} $ and
$\textrm{BR}(B_s \to \mu^+ \mu^-)_{\textrm{exp}} = (3.0 \pm 0.6) \times 10^{-9}$~\cite{1703.05747}.
This BR can also be affected by extra scalar operators $(\bar b P_X s)(\bar\mu P_Y \mu)$, so that it is sometimes omitted from
global BSM fits.

\section{Fits}\label{fit}

We divide the experimental data in two sets: `clean' and `hadronic sensitive':

\begin{itemize}

\item[\emph{i})]  The `clean' set includes the observables discussed in the previous section: $R_K$, $R_{K^*}$, to which one can add $\BR(B_s\to\mu^+\mu^-)$
given that it only provides constraints.\footnote{When using the {\sc Flavio}~\cite{Flavio} code, for consistency we include the observable $\BR(B_0\to\mu^+\mu^-)$, whose experimental error is correlated with the one on $\BR(B_s\to\mu^+\mu^-)$.} The `cleanness' of these observables refers to the SM prediction, in the presence of New Physics larger theoretical uncertainties are expected. We didn't include the $Q_4$ and $Q_5$ observables measured recently by the Belle collaboration \cite{1612.05014}.

\item[\emph{ii})] The `hadronic sensitive' set includes about 100 observables (summarized in the Appendix~\ref{appendixA}). 
This list includes the branching ratios of semi-leptonic $B$-meson decays as well as physical quantities extracted by the angular analysis of the decay products of the $B$-mesons. Concerning the hadronic sensitivity of the angular observables, the authors of \cite{1207.2753} argue that the optimised variables $P_i$ have reduced theoretical uncertainties.

\end{itemize}
The rationale is to first limit the analysis to the `clean' set of observables.
In this way one can draw solid conclusions without relying on large and partially uncontrolled effects.
This approach is aligned with the spirit of this paper, and can be extremely powerful, as already shown in section~\ref{sec:AnatomyOfRKStar}.
Furthermore, extracting from this reliable theoretical environment a BSM perspective could be of primary importance to set the stage for more complex analyses. In a second and third step we will estimate the effect of the `hadronic sensitive' observables and combine all observables in a global fit.

\subsection{Fit to the `clean' observables only}
\label{fit:clean}
The formul\ae{} summarized in the previous section allow us to  fit the clean observables.
We wrote a dedicated Flavour Anomaly Rate Tool code ({\sc Fart}).
For simplicity, in our $\chi^2$ fits we combine in quadrature the experimental errors on the two $R_{K^*}$ bins, using the higher error band when they are asymmetric.
We checked that our results do not change appreciably if a more precise treatment is used.

Let us start discussing the simplest case, in which we consider one-parameter fits to each NP operator in turn.
Apart from its simplicity, this hypothesis is motivated from a theoretical viewpoint, as it captures most of the relevant features of concrete models, as we shall discuss in detail in section~\ref{th}.
We show the corresponding results --- best-fit point, 1-$\sigma$ error,
and $\sqrt{\Delta\chi^2} \equiv \sqrt{\chi^2_{\rm SM} - \chi^{2}_{\rm best}}$ --- in the `clean' column in table~\ref{tab:FARTFit}.
In the upper part of the table we show the cases in which we allow new physics in the muon sector.
It is evident that the results of the fit match the discussion of section~\ref{sec:AnatomyOfRKStar}: the left-handed coefficient $C^{\rm BSM}_{b_L \mu_L}$  is favoured by the measured anomalies in $R_K$ and $R_{K^*}$, with a significance of about 4$\sigma$.
We can similarly discuss the hypothesis in which we allow for new physics in the electron sector, shown in the lower part of table~\ref{tab:FARTFit}.
Three cases --- $C^{\rm BSM}_{b_L e_L}$, $C^{\rm BSM}_{b_L e_R}$ and $C^{\rm BSM}_{b_R e_{R}}$ --- are equally favoured by the fit.
However, only the operator  $\mathcal{O}_{b_L e_L}$ involving left-handed quarks and electrons can explain the observed anomalies with an order one Wilson coefficient since it dominates the new-physics corrections to both $R_K$ and $R_{K^*}$, see Eqs~(\ref{eq:RKChiral},\ref{eq:RKStarChiral}).
As before, we find a statistical preference with respect to the SM case at the level of about 4-$\sigma$.
To simplify the comparison with the existing literature, we show in table~\ref{tab:FARTFit2} the results of $1$-parameter fits in the muon sector, this time in the vector-axial basis.

In conclusion, the piece of information that we learn from this simple fit is quite sharp: by restricting the analysis to the selected subset of  `clean' observables $R_K$, $R_{K^*}$ and ${\rm BR}(B_s \to \mu^+\mu^-)$, not much affected by large theoretical uncertainties, we find a  preference for the presence of new physics in the observed experimental anomalies in $B$ decays.
In particular, the analysis selects the existence of a new neutral current that couples left-handed $b$, $s$ quarks and left-handed muons/electrons  as the preferred option.

Effective four-fermions operators that couple left- or right-handed $b$, $s$ with right-handed electrons are also equally preferred at this level of the analysis, but they require
larger numerical values of their Wilson coefficients.

\begin{table}[pt]
\begin{center}
\begin{tabular}{||c||c|c|c||c|c|c||c|c|c||}
\hline\hline
  \multicolumn{10}{|c|}{\textbf{New physics in the muon sector}}   \\
\hline\hline
{\color{blue}{Wilson}} & \multicolumn{3}{c||}{{\color{blue}{Best-fit}}} & \multicolumn{3}{c||}{{\color{blue}{1-$\sigma$ range}}} &
 \multicolumn{3}{c||}{{\color{blue}{$\sqrt{\chi^2_{\rm SM} - \chi^2_{\rm best}}$}}}   \\ \cline{2-10}
{\color{blue}{coeff.}} &  {\color{giallo}{`clean'}}  & {\color{rossoc}{`HS'}} & {\color{verde}{all}}  & {\color{giallo}{`clean'}}  & {\color{rossoc}{`HS'}} & {\color{verde}{all}} & {\color{giallo}{`clean'}} & {\color{rossoc}{`HS'}} & {\color{verde}{all}}    \\ \hline
  \multirow{2}{*}{$C_{b_L \mu_L}^{\rm BSM}$} &  \multirow{2}{*}{$-1.27$} & \multirow{2}{*}{$-1.33$} &  \multirow{2}{*}{$-1.30$}  & $-0.94$ & $-1.01$  & $-1.07$ & \multirow{2}{*}{$4.1$} &  \multirow{2}{*}{$4.6$} & \multirow{2}{*}{$6.2$}   \\
  & & & & $-1.62$ & $-1.68$ & $-1.55$ & & & \\  \hline\hline
  \multirow{2}{*}{$C_{b_L \mu_R}^{\rm BSM}$} &  \multirow{2}{*}{$0.64$} & \multirow{2}{*}{$-0.73$} &  \multirow{2}{*}{$-0.30$}  & $1.17$ & $-0.40$  & $0.02$ & \multirow{2}{*}{$1.2$} &  \multirow{2}{*}{$2.1$} & \multirow{2}{*}{$0.9$}   \\
  & & & & $0.11$ & $-1.03$ & $-0.59$ & & & \\  \hline\hline
    \multirow{2}{*}{$C_{b_R \mu_L}^{\rm BSM}$} &  \multirow{2}{*}{$0.05$} & \multirow{2}{*}{$-0.20$} &  \multirow{2}{*}{$-0.14$}  & $0.33$ & $-0.04$  & $0.00$ & \multirow{2}{*}{$0.2$} &  \multirow{2}{*}{$1.3$} & \multirow{2}{*}{$1.0$}   \\
  & & & & $-0.23$ & $-0.29$ & $-0.25$ & & & \\  \hline\hline
      \multirow{2}{*}{$C_{b_R \mu_R}^{\rm BSM}$} &  \multirow{2}{*}{$-0.44$} & \multirow{2}{*}{$0.41$} &  \multirow{2}{*}{$0.27$}  & $0.08$ & $0.61$  & $0.48$ & \multirow{2}{*}{$0.8$} &  \multirow{2}{*}{$1.7$} & \multirow{2}{*}{$1.2$}   \\
  & & & & $-0.97$ & $0.18$ & $0.04$ & & & \\  \hline\hline
 \multicolumn{10}{|c|}{\textbf{New physics in the electron sector}}   \\
\hline\hline
{\color{blue}{Wilson}} & \multicolumn{3}{c||}{{\color{blue}{Best-fit}}} & \multicolumn{3}{c||}{{\color{blue}{1-$\sigma$ range}}} &
\multicolumn{3}{c||}{{\color{blue}{$\sqrt{\chi^2_{\rm SM} - \chi^2_{\rm best}}$}}}   \\ \cline{2-10}
{\color{blue}{coeff.}} &  {\color{giallo}{`clean'}}  & {\color{rossoc}{`HS'}} & {\color{verde}{all}}  & {\color{giallo}{`clean'}}  & {\color{rossoc}{`HS'}} & {\color{verde}{all}} & {\color{giallo}{`clean'}} & {\color{rossoc}{`HS'}} & {\color{verde}{all}}    \\ \hline
     \multirow{2}{*}{$C_{b_L e_L}^{\rm BSM}$} &  \multirow{2}{*}{$1.72$} & \multirow{2}{*}{$0.15$} &  \multirow{2}{*}{$0.99$}  & $2.31$ & $0.69$  & $1.30$ & \multirow{2}{*}{$4.1$} &  \multirow{2}{*}{$0.3$} & \multirow{2}{*}{$3.5$}   \\
 & & & & $1.21$ & $-0.39$ & $0.70$ & & & \\  \hline\hline
     \multirow{2}{*}{$C_{b_L e_R}^{\rm BSM}$} &  \multirow{2}{*}{$-5.15$} & \multirow{2}{*}{$-1.70$} &  \multirow{2}{*}{$-3.46$}  & $-4.23$ & $0.33$  & $-2.81$ & \multirow{2}{*}{$4.3$} &  \multirow{2}{*}{$0.9$} & \multirow{2}{*}{$3.6$}   \\
 & & & & $-6.10$ & $-2.83$ & $-4.05$ & & & \\  \hline\hline
      \multirow{2}{*}{$C_{b_R e_L}^{\rm BSM}$} &  \multirow{2}{*}{$0.085$} & \multirow{2}{*}{$-0.51$} &  \multirow{2}{*}{$0.02$}  & $0.39$ & $0.29$  & $0.30$ & \multirow{2}{*}{$0.3$} &  \multirow{2}{*}{$0.7$} & \multirow{2}{*}{$0.1$}   \\
 & & & & $-0.21$ & $-1.55$ & $-0.25$ & & & \\  \hline\hline
      \multirow{2}{*}{$C_{b_R e_R}^{\rm BSM}$} &  \multirow{2}{*}{$-5.60$} & \multirow{2}{*}{$2.10$} &  \multirow{2}{*}{$-3.63$}  & $-4.66$ & $3.52$  & $-2.65$ & \multirow{2}{*}{$4.2$} &  \multirow{2}{*}{$0.5$} & \multirow{2}{*}{$2.5$}   \\
 & & & & $-6.56$ & $-2.70$ & $-4.43$ & & & \\  \hline\hline
  \end{tabular}
\end{center}
\caption{\em Best fits assuming a single chiral operator at a time, and fitting only the `clean' $R_K$, $R_{K^*}$, and ${\rm BR}(B_s \to \mu^+\mu^-)$, or only the `Hadronic Sensitive' observables  (denoted by `HS' in the table) as discussed in the text, or combining them in a global fit.
The full list of observable can be find in appendix~\ref{appendixA}.
\label{tab:FARTFit}}
\end{table}%

Needless to say, this conclusion, although already very significant,
must be supported by the result of a more complete analysis that accounts for all the other observables related to $B$ decays, and not included
in the `clean' set used in this section. We shall return to this point in section~\ref{sec:Dirty}.

\begin{table}[pt]
\begin{center}
\begin{tabular}{||c||c|c|c||c|c|c||c|c|c||}
\hline\hline
  \multicolumn{10}{|c|}{\textbf{New physics in the muon sector (Vector Axial basis)}}   \\
\hline\hline
{\color{blue}{Wilson}} & \multicolumn{3}{c||}{{\color{blue}{Best-fit}}} & \multicolumn{3}{c||}{{\color{blue}{1-$\sigma$ range}}} &
 \multicolumn{3}{c||}{{\color{blue}{$\sqrt{\chi^2_{\rm SM} - \chi^2_{\rm best}}$}}}   \\ \cline{2-10}
{\color{blue}{coeff.}} &  {\color{giallo}{`clean'}}  & {\color{rossoc}{`HS'}} & {\color{verde}{all}}  & {\color{giallo}{`clean'}}  & {\color{rossoc}{`HS'}} & {\color{verde}{all}} & {\color{giallo}{`clean'}} & {\color{rossoc}{`HS'}} & {\color{verde}{all}}    \\ \hline
  \multirow{2}{*}{$C_{9,\,\mu}^{\rm BSM}$} &  \multirow{2}{*}{$-1.51$} & \multirow{2}{*}{$-1.15$} &  \multirow{2}{*}{$-1.19$}  & $-1.05$ & $-0.98$  & $-1.04$ & \multirow{2}{*}{$3.9$} &  \multirow{2}{*}{$5.5$} & \multirow{2}{*}{$6.7$}   \\
  & & & & $-2.08$ & $-1.31$ & $-1.35$ & & & \\  \hline\hline
  \multirow{2}{*}{$C_{10,\,\mu}^{\rm BSM}$} &  \multirow{2}{*}{$0.97$} & \multirow{2}{*}{$0.48$} &  \multirow{2}{*}{$0.66$}  & $1.28$ & $0.69$ & $0.83$ & \multirow{2}{*}{$3.8$} &  \multirow{2}{*}{$2.4$} & \multirow{2}{*}{$4.3$}   \\
  & & & & $0.69$ & $0.28$ & $0.50$ & & & \\  \hline\hline
    \multirow{2}{*}{$C_{9,\,\mu}^{\prime{\rm BSM}}$} &  \multirow{2}{*}{$-0.08$} & \multirow{2}{*}{$-0.24$} &  \multirow{2}{*}{$-0.22$}  & $0.20$ & $-0.15$  & $-0.14$ & \multirow{2}{*}{$0.3$} &  \multirow{2}{*}{$1.7$} & \multirow{2}{*}{$1.6$}   \\
  & & & & $-0.37$ & $-0.36$ & $-0.33$ & & & \\  \hline\hline
      \multirow{2}{*}{$C_{10,\,\mu}^{\prime{\rm BSM}}$} &  \multirow{2}{*}{$-0.11$} & \multirow{2}{*}{$0.10$} &  \multirow{2}{*}{$0.07$}  & $0.11$ & $0.19$  & $0.15$ & \multirow{2}{*}{$0.5$} &  \multirow{2}{*}{$1.2$} & \multirow{2}{*}{$0.9$}   \\
  & & & & $-0.34$ & $0.01$ & $-0.01$ & & & \\  \hline\hline
   \end{tabular}
\end{center}
\caption{\em Same as table~\ref{tab:FARTFit}, but in the vector-axial basis.
\label{tab:FARTFit2}}
\end{table}%

\begin{figure}[!htb!]
\begin{center}
\includegraphics[width=0.99\textwidth]{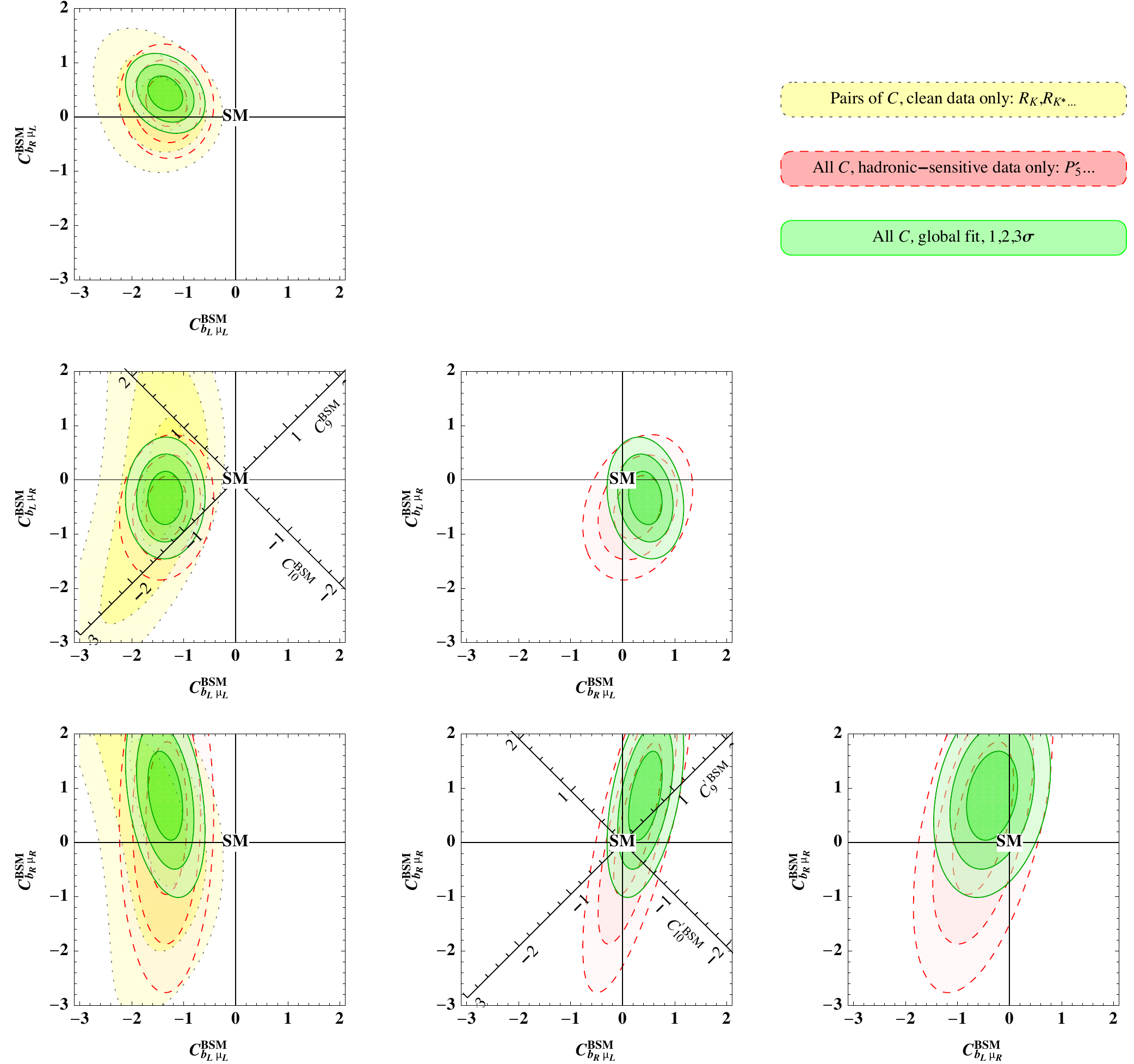}
\caption{\em Fit to the new-physics contribution to the coefficients of the
4 muon operators $(\bar b \gamma_\mu  P_X s)(\bar\mu \gamma_\mu P_Y \mu)$, showing the $1,2,3\sigma$ contours.
The yellow regions with dotted contours show the best fit to the  `clean' observables only; due to the scarcity of data, in each plot we turn on only the two coefficients indicated on its axes.
The red regions with dashed contours show the best global fit to the `hadronic sensitive' observables only, according to one estimate of their theoretical uncertainties; in this fit, we turn on all 4 muon operators at the same time and, in each plot, we marginalise over the coefficients not shown in the plot.
The green regions show the global fit, again  turning on all 4 muon operators at the same time.
In figure\fig{fitglobalmue} we  turn on the extra 4 electron operators too.
\label{fig:fitglobalmu}}
\end{center}
\end{figure}

Before moving to the fit with the `hadronic sensitive' observables, we perform several two-parameter fits using only `clean' observables.
We show our results in figure\fig{fitglobalmu}.
Allowing for new physics in muons only, the combined best-fit regions are shown as yellow contours.
Since there are few `clean' observables, we turn on only two new-physics coefficients in each plot, as indicated on the axes.
We also show, as rotated axes, the usual $C_9$ and $C_{10}$ coefficients.
We see that the key implications mentioned in section~\ref{RK*an} are confirmed by this fit, although here wider regions in parameter space are allowed.
In the upper plot of figure \fig{fitglobalmu} we show the results for new physics in the operators involving left-handed muons, $C_{b_L\mu_L}$ and $C_{b_R\mu_R}$: both coefficients are fixed by the `clean' data.
Operators involving right-handed muons, on the other hand, do not lead to good fits.
A good fit is obtained by turning on only $C_{b_L\mu_L}$, although uncertainties do not yet allow to draw sharp conclusions.

We conclude this section with a comment on the size of the theoretical uncertainties in the presence of New Physics.
While there is a consensus on the small error of the Standard Model predictions, in the presence of New Physics the ``clean'' observables have a larger theoretical error, barring the special case where new physics violate flavour universality while maintaining the same chiral structure of the SM
(mostly $LL$ at large enough $q^2$).
As shown in figure~\ref{fig:RKAnatomy}, away from the Standard Model our errors are still of a few percent, in agreement with ref.~\cite{1704.05446, 1704.05435}.
However, other groups~\cite{1704.05340} find a much larger theoretical error in the presence of New Physics, due to a more conservative treatment of the form factor uncertainties.\footnote{We thank Joaquim Matias for enlightening discussions about this point.}
Therefore, we warn the reader that the statistical significance quoted in our fits may be smaller with a different treatment of the error.

We didn't take into account another important source of error: QED radiative corrections, calculated in ref.~\cite{1605.07633}.
These are of the same order or larger than the hadronic uncertainties on $R_K$, $R_{K*}$ in the Standard Model as predicted by {\sc Flavio}.
We did the exercise of inflating our hadronic error by a factor of 3, finding indeed a larger error away from the Standard Model, but still of the same order of the QED corrections.

\begin{figure}[tp]
\begin{center}
$$\includegraphics[width=0.4\textwidth]{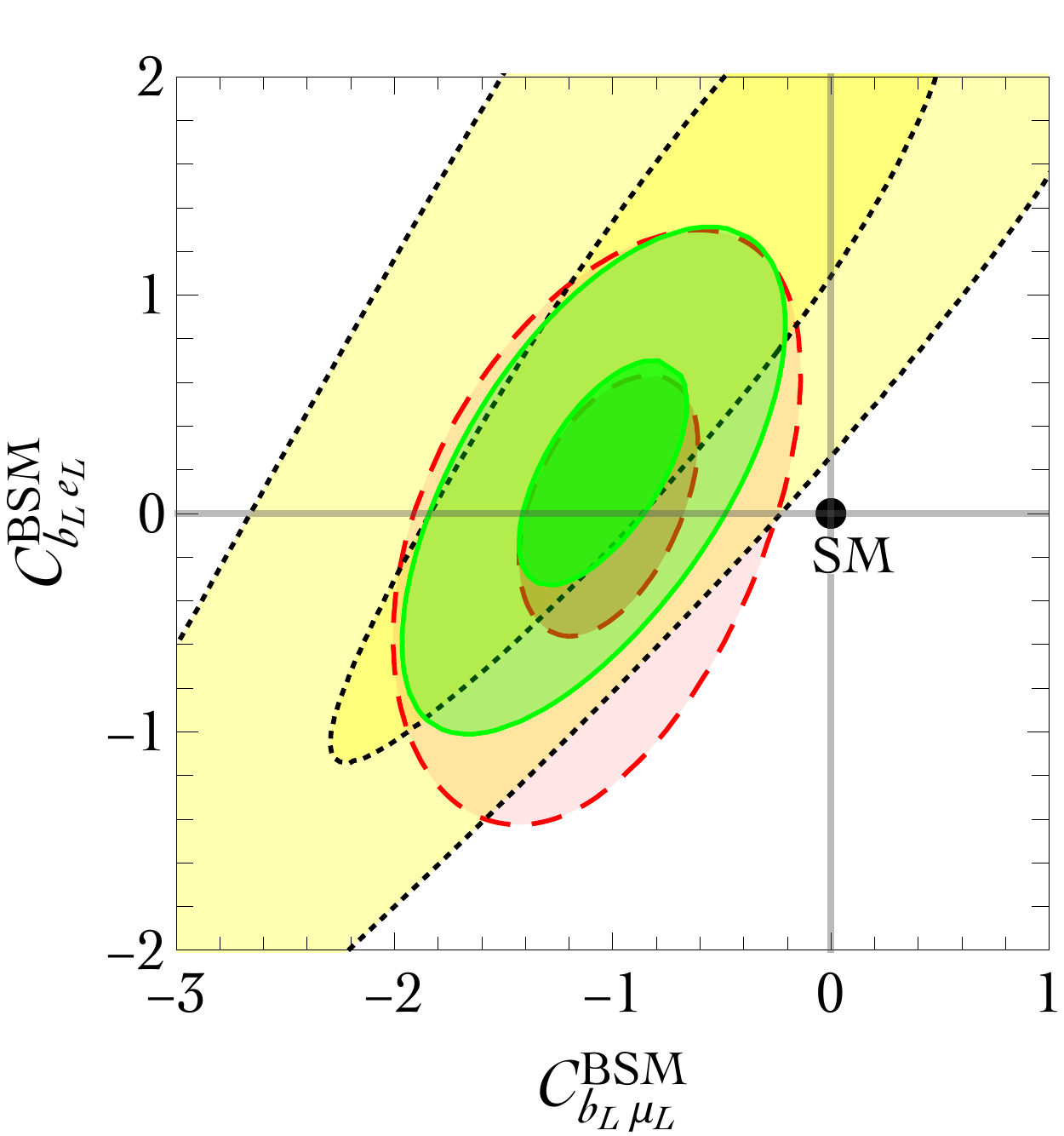}\qquad\includegraphics[width=0.4\textwidth]{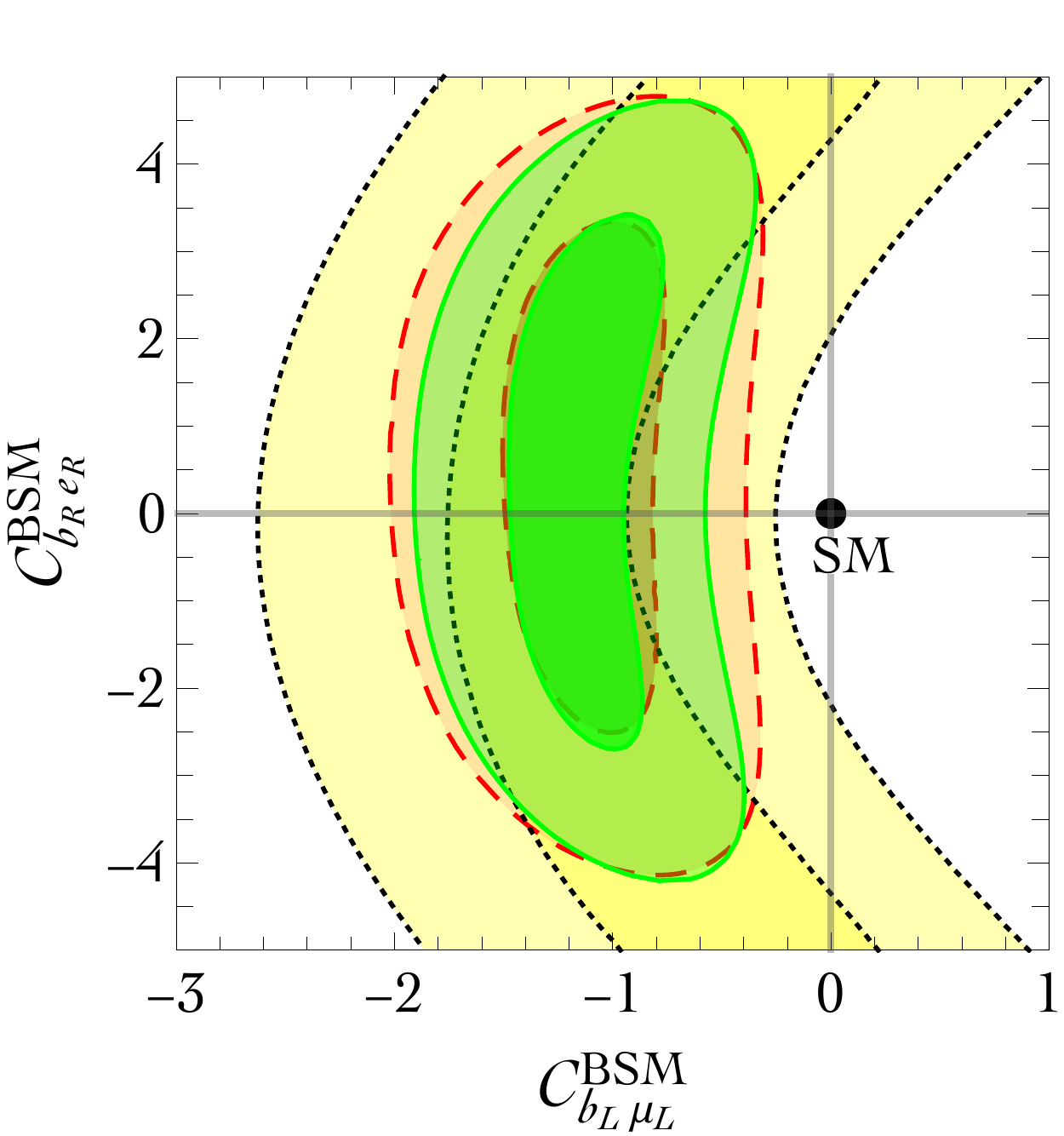}$$
\caption{\em Fits allowing one operator involving muons (horizontal axis) and one involving electrons (vertical axis):
left-handed in the left panel, and right-handed in the right panel.
Regions and contours have the same meaning as in fig.\fig{fitglobalmu}: `clean' data can be fitted by an anomaly in muons or electrons;
`hadronic sensitive' data favour an anomaly in muons.
\label{fig:fit2mue}}
\end{center}
\end{figure}

\subsection{Fit to the `hadronic sensitive' observables}\label{sec:Dirty}
In order to perform a global fit using the `hadronic sensitive' observables we use the public code {\sc Flavio}~\cite{Flavio}.

Theoretical uncertainties are dominant, and it is difficult to quantify them.
We first take theoretical uncertainties into account using the `FastFit' method in the {\sc Flavio} code with the addition of all the included nuisance parameters.
With this choice, the SM is disfavoured at about $5\sigma$ level.

Given that most `hadronic sensitive' observables involve muons (detailed measurements are much more difficult with electrons),
we present a simple $\chi^2$ of the 4 Wilson coefficients involving muons.
This is a simple useful summary of the full analysis.
In this approximation, the `hadronic sensitive' observables determine the 4 muon Wilson coefficients as\footnote{In  general, within the Gaussian approximation, the mean values $\mu_i$, the errors $\sigma_i$ and the correlation matrix
$\rho_{ij}$ determine the $\chi^2$ as
$\chi^2 =\sum_{i,j} (C_i - \mu_i) (\sigma^2)^{-1}_{ij}  (C_j - \mu_j)$, {where} $ (\sigma^2)_{ij} = \sigma_i \rho_{ij} \sigma_j\ $.}
\beq\label{eq:Cmudirty}
\begin{array}{rcl}
C_{b_L\mu _L}^{\text{BSM}} &=& -1.33 \pm 0.26\\
C_{b_R\mu _L}^{\text{BSM}} &=& +0.29 \pm 0.31\\
C_{b_L\mu _R}^{\text{BSM}} &=& -0.51 \pm 0.39\\
C_{b_R\mu _R}^{\text{BSM}} &=& +0.45 \pm 0.93\\
\end{array}\qquad\hbox{with}\qquad\rho =
\left(
\begin{array}{cccc}
 1 & -0.07 & 0.13 & 0.03 \\
 -0.07 & 1 & 0.25 & 0.74 \\
 0.13 & 0.25 & 1 & 0.50 \\
 0.03 & 0.74 & 0.50 & 1 \\
\end{array}
\right)
.\eeq
The uncertainties can be rescaled by factors of ${\cal O}(1)$,
if one believes that theoretical uncertainties should be larger or smaller than those adopted here.

The global fit of `hadronic sensitive' observables to new physics in the 4 muon coefficients is also shown as red regions in figure\fig{fitglobalmu}.
The important message is apparent both from the figure and from eq.\eq{Cmudirty}: {\em `hadronic sensitive' observables favour a deviation from the SM
in the same direction as the `clean' observables}, i.e. a negative contribution $C^{\rm BSM}_{b_L\mu_L} \sim -1$ to the Wilson coefficient
involving left-handed quarks and muons.
`Clean' observables and `hadronic sensitive' observables --- whatever  their uncertainty is --- look consistent and favour independently the same pattern
of deviations from the SM.

\subsection{Global fit}
We are now ready to combine `clean' and `hadronic sensitive' observables in a global fit, using both the {\sc Flavio} and {\sc Fart} codes.
The result is shown as green regions in figure\fig{fitglobalmu}, assuming that new physics affects muons only.
The global fit favours a deviation in the SM in $C^{\rm BSM}_{b_L\mu_L}$, and provides bounds on the other new-physics  coefficients.
Using the Gaussian approximation for the likelihood of the muon coefficients, the global fit is summarized as
\beq\label{eq:muold}
\begin{array}{rcl}
C_{b_L\mu _L}^{\text{BSM}} &=& -1.35 \pm 0.22\\
C_{b_R\mu _L}^{\text{BSM}} &=& +0.44 \pm 0.21\\
C_{b_L\mu _R}^{\text{BSM}} &=& -0.33 \pm 0.33\\
C_{b_R\mu _R}^{\text{BSM}} &=& +0.86 \pm 0.54\\
\end{array}\qquad\hbox{with}\qquad\rho =
\left(
\begin{array}{cccc}
 1 & -0.26 & 0.02 & -0.33 \\
 -0.26 & 1 & -0.17 & 0.47 \\
 0.02 & -0.17 & 1 & 0.25 \\
 -0.33 & 0.47 & 0.25 & 1 \\
\end{array}
\right)
.\eeq

An anomaly in muons is strongly preferred to an anomaly in electrons, if we adopt
the default estimate of the theoretical uncertainties by FLAVIO.
This is for example shown in fig.\fig{fit2mue}, where we allow for a single
 operator involving muons and a single operator involving electrons.

\begin{figure}[tp]
\begin{center}
\includegraphics[width=0.99\textwidth]{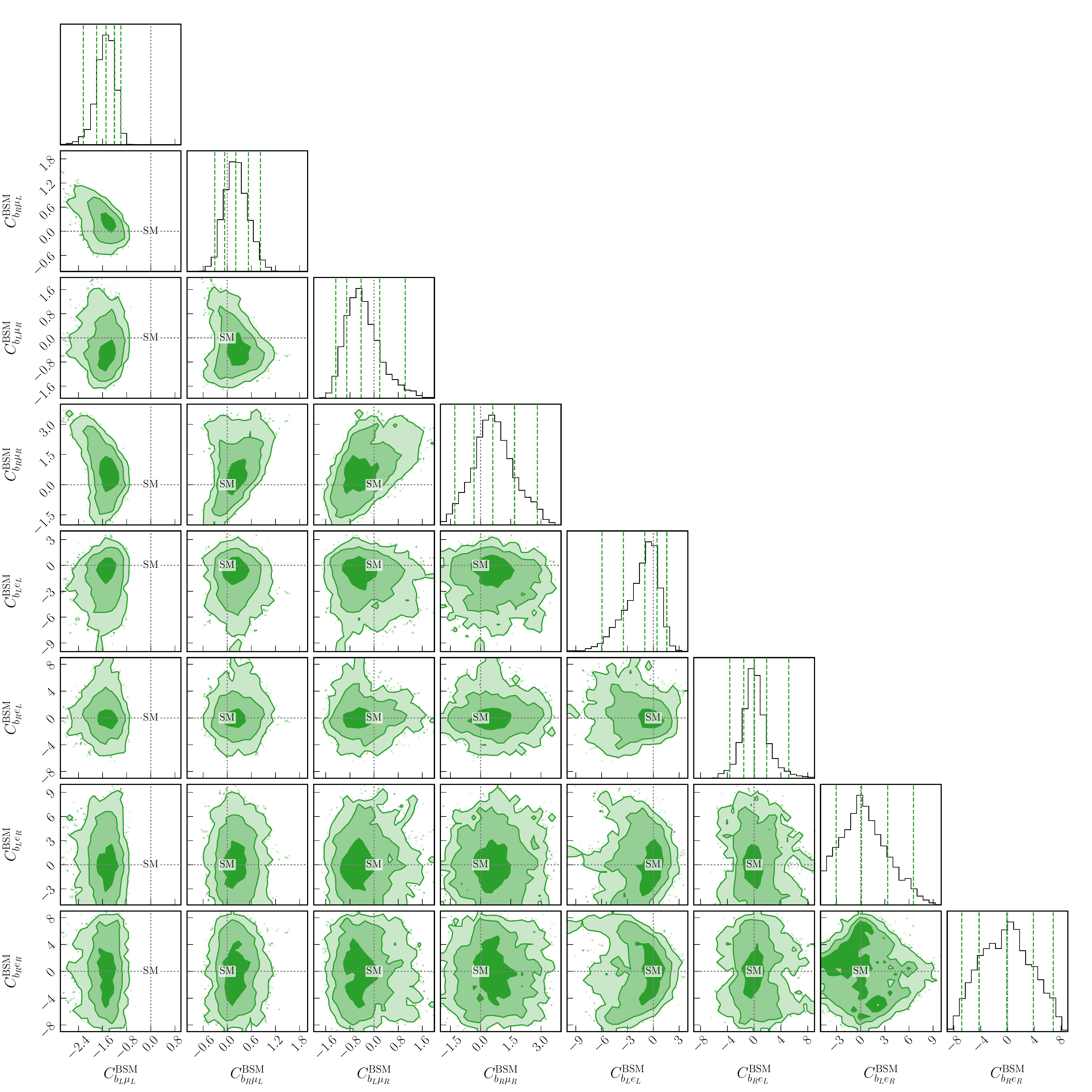}
\caption{\em Global fit to the 8 Wilson coefficients in muons and electrons, combining `clean' and `hadronic sensitive' data.
\label{fig:fitglobalmue}}
\end{center}
\end{figure}

In view of this preference, and given the scarcity of data in the electron sector, we avoid presenting a global fit of new physics in electrons only.
We instead perform a global combined fit for the muon and electron coefficients (which should be interpreted with caution, given that `hadronic sensitive' observables are dominated by theoretical uncertainties).
We find the result shown in figure\fig{fitglobalmue}, which confirms
that --- while electrons can be affected by new physics --- `hadronic sensitive' data favour an anomaly in muons.

The latter result has been obtained by a global Bayesian fit to the observables listed in tables~\ref{tab:ang_obs}, \ref{tab:br_obs} in addition to the clean observables.
We used the \textsc{Flavio} code to calculate the likelihood, and we sampled the posterior using the \textsc{Emcee} code~\cite{1202.3665}, assuming for the 8 Wilson coefficients (at the scale $160 \GeV$) a flat prior between $-10$ and $10$.
In this global fit, we choose to marginalize over $25$ nuisance parameters only, to keep computational times within reasonable limits.
The nuisances (form factors related to $B$ decays) are selected in the following way.
For each observable, we define theoretical uncertainties due to changing each nuisance within its uncertainty, keeping the others fixed at their central values.
Then, we choose to marginalize only over the parameters which give a theoretical uncertainty larger than the experimental error on the observable.

\begin{figure}[t]
\begin{center}
\includegraphics[width=0.99\textwidth]{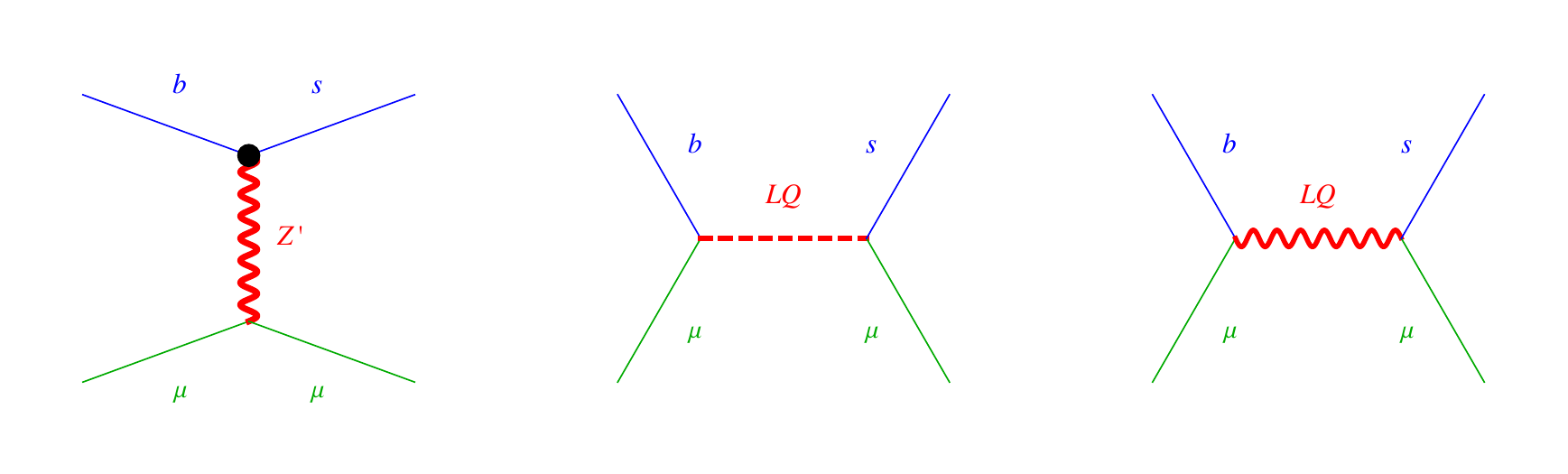}
\caption{\em Particles that can mediate $R_K$ at tree level: a $Z'$ or a lepto-quark, scalar or vector.
\label{fig:FeynZLQ}}
\end{center}
\end{figure}

\section{Theoretical interpretations}\label{th}
We now discuss different theoretical interpretations that can accommodate the flavour anomalies.
We start with the observation that an effective $(\bar s \gamma_\mu P_X b)(\bar \ell \gamma_\mu P_Y \ell)$ interaction can be mediated at tree level by two kinds of particle: a $Z'$ or a leptoquark.
Higher-order  induced mechanisms are also possible.
These models tend to generate related operators
\beq c_{b_L b_L} (\bar s \gamma_\mu P_L b)^2+
c_{\mu_L\nu_\mu}
( \bar{\mu}\gamma^{\mu} P_L \mu )(\bar{\nu}_\mu \gamma_\mu P_L\nu_\mu) \ ,
\eeq
and therefore one needs to consider the associated experimental constraints.  The first operator affects  $B_s$ mass mixing for which the relative measurements,  together with  CKM fits, imply
$c^{\rm BSM}_{b_L b_L} = (-0.09 \pm 0.08)/(110\TeV)^2$ , i.e. the bound $|c^{\rm BSM}_{b_L b_L}|< 1/ (210\TeV)^2$~\cite{1408.4097,1608.07832}.
The second operator is constrained by CCFR data on the neutrino trident cross section, yielding the weaker bound
$|c^{\rm BSM}_{\mu_L\nu_\mu}|< 1/(490\GeV)^2$ at 95\% C.L.~\cite{1406.2332}.
Furthermore, new physics that affects muons can contribute to the anomalous magnetic moment of the muon.
Experiments found hints of a possible deviation from the Standard Model with $\Delta a _\mu = (24\pm 9)\cdot 10^{-10}$~\cite{hep-ex/0602035}.

\subsection{Models with an extra $Z'$}
Models featuring extra $Z'$  to explain the anomalies are very popular, see the partial list of references~\cite{1310.1082,1311.6729,1403.1269,
1501.00993,
1503.03477,
1503.03865,
1505.03079,
1506.01705,
1508.07009,
1509.01249,1510.07658,
1512.08500,
1601.07328,
1602.00881,
1604.03088,
1608.01349,1608.02362,1611.03507,1702.08666,1703.06019}.
Typically these models contain a $Z'$ with mass $M_{Z'}$  savagely coupled to
\beq [g_{bs} (\bar s \gamma_\mu P_L b) +\hbox{h.c.}] + g_{\mu_L} (\bar\mu \gamma_\mu P_L \mu) \ .\eeq
The model can reproduce the flavour anomalies with
$c_{b_L \mu_L} =- g_{bs} g_{\mu_L}/M_{Z'}^2$ as illustrated in figure\fig{FeynZLQ}a.
At the same time the $Z'$ contributes to the ${B_s}$ mass mixing with
$c_{b_L b_L} =- g_{bs}^2/2 M_{Z'}^2$.
The bound from $\Delta M_{B_s}$ can be satisfied by requiring a large enough $g_{\mu_L}$ in order to reproduce the $b\to s\ell^+\ell^-$ anomalies.
 Left-handed leptons are unified in a $\SU(2)_L$ doublet $L = (\nu_L,\ell_L)$,
such that also the neutrino operator $c_{\mu_L \nu_\mu} =- g_{\mu_L}^2/M_{Z'}^2$ is generated. However the latter does not yield a strong constraint on $g_{\mu_L}$.

Another possibility is for the $Z'$ to couple  to the  3-rd generation left-handed quarks with coupling $g_t$ and to lighter left-handed quarks with coupling $g_q$.
The coupling $g_{bs}$ arises as $g_{bs} = (g_t-g_q) (U_{Q_d})_{ts}$
after performing a flavour rotation $U_{Q_d}$ among left-handed down quarks
to their mass-eigenstate basis.
The  matrix element $(U_{Q_d})_{ts}$ is presumably not much larger than $V_{ts}$ and possibly equal to it,
if the CKM matrix $V = U_{Q_u} U_{Q_d}^\dagger$ is dominated by the rotation among left-handed down quarks,
rather than by the rotation $U_{Q_u}$ among left-handed up quarks.

Unless $g_q=0$, the parameter space of the $Z'$ model gets severely constrained by combining perturbative bounds on $g_{\mu_L}$.  In addition the LHC bounds on $pp\to Z'\to \mu\bar\mu$  can be relaxed by introducing extra features, such as a $Z'$ branching ratio into invisible DM particles~\cite{1511.07447}.

A characteristic feature of $Z'$ models is that they can mediate effective operators involving different chiralities.
In fact, gauge-anomaly cancellations also induce multiple chiralities:
for example a $Z'$ coupled to $L_\mu- L_\tau$ is anomaly free~\cite{1403.1269},
where the  $L_e$ contribution is avoided because LEP put  strong  constraints on 4-electron operators.
The chiralities involved in the $b\to s\ell^+\ell^-$ anomalies can be determined trough
more precise measurements of  `clean' observables such as $R_K$ and $R_{K^*}$.

\subsection{Models with lepto-quarks}\label{LQ}
The anomalous effects in $b \to s \ell^+ \ell^-$ transitions might be due to the exchange of a Lepto-Quark (LQ),
namely a boson that couples to a lepton and a quark.
Concerning lepton flavour, in general a LQ can couple to both muons and electrons.
However, simultaneous sizeable couplings of a LQ to electrons and muons generates lepton flavour violation which is severely constrained by the time-honoured radiative decay  $\mu \to e \gamma$. For this reason one typically assumes that
LQs couple to either electrons or muons.
(Here sizeable means an effect which has an impact  on the anomalous observables).
The coupling to muons allows to fit the anomalies in $b\to s\mu^+\mu^-$ distributions, as well as the $R_K$ and $R_{K^*}$ $\mu/e$  ratios.

The gauge quantum numbers of \textit{scalar} LQs  select a specific chirality
of the SM fermions involved in the new Yukawa couplings,
and thereby  generate a unique characteristic operator
in the effective Lagrangian in the chiral basis of eq.\eq{Leff}, as illustrated in figure\fig{FeynZLQ}b.
The correspondence is given by
\beq
\begin{array}{ccc}
\hbox{Coefficient} & \hbox{Lepto-Quark} & \hbox{Yukawa couplings}\\ \hline
C_{b_L \ell_L} &     S_3 \sim ({\bar{3}},{3},1/3)  &  y \,  Q L \, S_3 + y' \, QQ \, S^{\dagger}_3 +\hbox{h.c.} \\
C_{b_L \ell_R} &     R_2 \sim (3,{2},7/6) &  y \, UL \, R_2 + y' \, QE  \, R^{\dagger}_2 +\hbox{h.c.}  \\
C_{b_R \ell_L} &     \tilde{R}_2 \sim (3,{2},1/6) & y \,DL \, \tilde{R}_2 + \hbox{h.c.}  \\
C_{b_R \ell_R} &     \tilde{S}_1 \sim  ({\bar{3}},{1},4/3)  & y \, DE \, \tilde{S}_1 + y' \, UU \, \tilde{S}^{\dagger}_1 +\hbox{h.c.}
\end{array}
\eeq
where $\ell$ can be either an electron or a muon.
In parentheses we report the $\SU(3)\times \SU(2)_L\times {\rm U}(1)_Y$ gauge quantum numbers, and we follow the notations and conventions from ref.~\cite{1603.04993} for LQ names.
$Q,L$ ($U,D,E$) denote the left-handed (right-handed) SM quarks and leptons.

Given that each LQ mediates effective operators with a given chirality,
we can draw conclusions from our one parameter fits of the $b\to s\ell^+\ell^-$ anomalies of table~\ref{tab:FARTFit}.
Assuming new physics in the muon sector,  the measurement of $R_{K^*}$ selects a unique scalar lepto-quark: $S_3 \sim ({\bar{3}},{3},1/3)$, which is a
triplet under $\SU(2)_L$.
It is remarkable that this is obtained with just the information coming from `clean' observables while the inclusion of the remaining observables (with our specified treatment of the errors) reinforces this hypothesis. The explanation of the anomalies in terms of $S_3$ has been firstly proposed after the measurement of $R_K$ in ref.~\cite{1408.1627} switching on only those couplings needed to reproduce the effect. In ref.~\cite{1412.1791} the LQ has been identified as a pseudo-Goldstone boson associated to the breaking of a global symmetry of a new strongly coupled sector \cite{0910.1789}. In ref.~\cite{0910.1789,1412.1791} it has also been suggested that a rationale for the size of the various flavour couplings could be dictated by the mechanism of partial compositeness \cite{Kaplan:1991dc}. Another motivated pattern of couplings has been suggested in ref.~\cite{1503.01084} using flavour symmetry. Also ref.~\cite{1703.09226} makes use of $S_3$ as mediator of the $b \to s \mu^+ \mu^-$ transition.

A potential issue with $S_3$ is the  danger of extra renormalizable couplings with di-quarks (denoted collectively by $y'$ in the Lagrangians above) which may induce proton decay. Baryon number conservation has to be invoked to avoid this issue.
Motivated by this, in ref.~\cite{1501.03494,1503.09024}, the LQ $\tilde{R}_2$  (which respects the global symmetry U$(1)_B$ accidentally at the renormalizable level) has been considered leading to the prediction $R_{K^*}>1$, which is now disfavoured by the LHCb data.
The other two options $\tilde{S}_1$ and $R_2$ were already disfavoured after the measurement of $R_K$~\cite{1408.1627,1501.05193}.

\medskip

The situation is different if LQs  couple to electrons, rather than to muons, such that only the anomalies in the  `clean' observables can be reproduced.
`Clean' observables can be reproduced by all chiralities, with the only exclusion of  $C_{b_R \ell_L}$, which is mediated by the $\tilde{R}_2$ LQ. From the fit, we notice that the $\tilde{S}_1$ and $R_2$ LQs can only fit the anomalies by giving a large contribution to the Wilson coefficients, comparable to the SM contributions: this happens  because these LQs couple to right handed electrons, with little interference with the SM.
One the other hand, $S_3$ couples to left-handed leptons, such that the sizeable interference with the SM allows to reproduce the observed anomalies with a smaller new physics component.

\begin{table}[t]
$$\begin{array}{c|c|c|c|c|c}
 & \textrm{Spin} & \textrm{Quantum} & \textrm{Clean observables} & \textrm{Clean observables} & \textrm{All} \\
 &                       &  \textrm{Number}  & \textrm{new physics in } e & \textrm{new physics in } \mu & \textrm{observables} \\
\hline
\hline
S_3           & 0 &  ({\bar{3}},{3},1/3)          & \checkmark  & \checkmark &  \checkmark \\
R_2           & 0 &  (3,{2},7/6)                     & \checkmark & &  \\
\tilde{R}_2 & 0 &  (3,{2},1/6)                     &  & &  \\
\tilde{S}_1 & 0 & ({\bar{3}},{1},4/3)           & \checkmark & &  \\
\hline
U_3           & 1 & (3,3,2/3)                        & \checkmark  & \checkmark &  \checkmark  \\
V_2           & 1 &  (\overline{3},2,5/6)       & \checkmark & &  \\
U_1           & 1 & (\overline{3},1,2/3)        & \checkmark  & \checkmark &  \checkmark  \\
\end{array}
$$
\caption{\em Which lepto-quarks can reproduce which $b\to s \ell^+\ell^-$ anomalies.\label{tab:LQtable}}
\end{table}

\medskip

We briefly comment on the possible interpretation of a LQ as a supersymmetric particle in the MSSM.
The only sparticle with the same gauge quantum numbers as a LQ is the left-handed squark $\tilde{Q}\sim \tilde{R}_2$.
However, even if it has $R$-parity violating interactions, this LQ gives the wrong correlation between $R_K$ and $R_{K^*}$, disfavouring the supersymmetric interpretation of the anomalies.

\medskip

We move now to the discussion of the exchange of  \textit{vector} LQs at tree level,  illustrated in figure\fig{FeynZLQ}c.
There are 3 cases: $U_3 \sim (3,3,2/3)$, $V_2 \sim (\overline{3},2,5/6)$ and $U_1 \sim (\overline{3},1,2/3)$.
Their relevant interactions are:
\begin{eqnsystem}{sys:VLQL}
\Lag_{U_3} &=& y  \, \bar{Q} \gamma_{\mu} L \, U^{\mu}_3 +\hbox{h.c.}  \\
\Lag_{V_2} &=& y  \,  \bar D  \gamma_{\mu} L \, V_2^{\mu} + y'  \,   \bar{Q} \gamma_{\mu} E \, V_2^{\mu} +  y''  \,   \bar{Q} \gamma_{\mu} U \, V_2^{\dagger \mu} + \hbox{h.c.}\\
\Lag_{U_1} &=& y  \,  \bar{Q}  \gamma_{\mu} L \, U_1^{\mu} + y_2  \,   \bar{D} \gamma_{\mu} E \, U_1^{\mu} + \hbox{h.c.}
\end{eqnsystem}
The vector LQ $V_2$ and $U_1$ can contribute to the anomalous observables trough multiple chiral structures.
In general, if both $y$ and $y'$ are sizeable, dangerous scalar operators may be generated.
If one of the two couplings dominates, we can again restrict to our one parameter fit, with the following correspondence:
$C_{b_L \ell_L}$ can be generated  by $U_3$;
$C_{b_L \ell_R}$ or $C_{b_R \ell_L} $ can be generated  by $V_2$;
$C_{b_L \ell_L}$ or $C_{b_R \ell_R} $ can be generated  by $U_1$.

Similar phenomenological considerations to explain the $B$-meson anomalies as in the case of the scalar LQ apply, we summarise the relevant options in table~\ref{tab:LQtable}.

Models featuring vector LQs models in order to explain the flavour anomalies appeared recently in the literature \cite{1505.05164,1512.01560,1609.04367,1611.04930}, typically as new composite states. The presence of these states signals that the theory in isolation is non-renormalizable, meaning that loop effects of the vectors are UV divergent, for a recent re-discussion see ref.~\cite{1607.07621}. Naive dimensional analysis shows that one-loop contributions to physics observables such $\Delta M_{B_s}$ might be problematic. A careful study of this topic is a model dependent issue and it requires extra information on the UV embedding of the LQ  in a complete theory.

\begin{figure}[t]
\begin{center}
\includegraphics[width=0.99\textwidth]{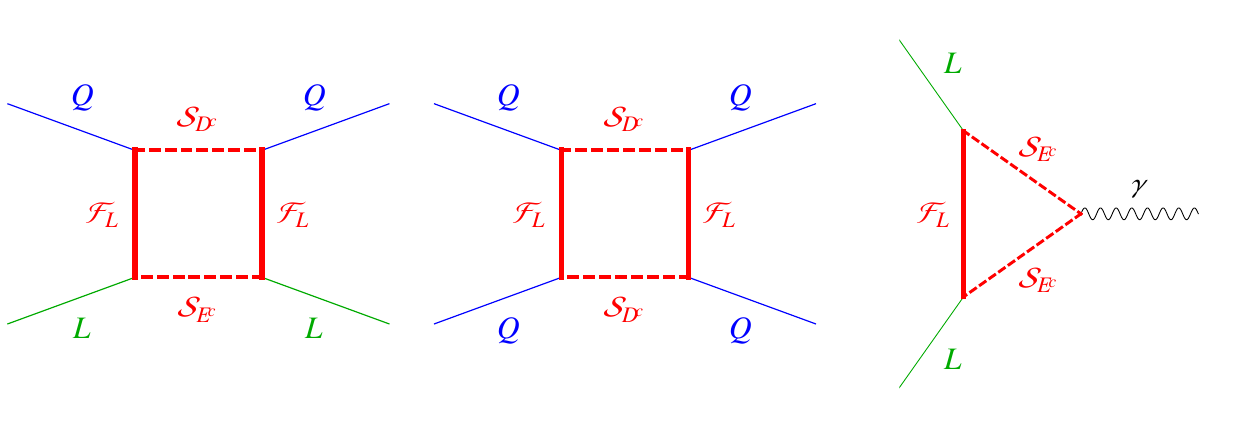}
\caption{\em Feynman diagrams contributing to $R_K$, $\Delta M_{B_s}$ and the muon $g-2$ in models with extra fermions ${\cal F}$
and extra scalars ${\cal S}$.
In Fundamental Composite Higgs models these diagrams will be dressed by further new composite dynamic contributions.
\label{fig:FeynRK}}
\end{center}
\end{figure}

\subsection{Models with loop mediators}
The $R_K$ anomaly can be reproduced by one loop diagrams involving new scalars ${\cal S}$ and new fermions  ${\cal F}$
with Yukawa couplings to SM fermions that allow for the Feynman diagram on the left in figure\fig{FeynRK}~\cite{1509.05020,1608.07832} -- see also ref.~\cite{1507.06660}.
In this particular example, one generates an operator involving  left-handed SM quarks and leptons, denoted  respectively by $Q$ and $L$.
The needed extra Yukawa coupling to the muon must be large, $y_L \sim 1.5$.  This also explains why the MSSM does not allow for an explanation of the ${R_K}, R_{K^*}$ anomalies: a possibile box diagram containing Winos and sleptons predicts $y_L \sim g_2 \ll 1.5$, where $g_2$ is the $\SU(2)_L$ gauge coupling.

In section~\ref{FPC} we will consider renomalizable models of composite dynamics featuring extra elementary scalars, where we will show
that the  extra particles ${\cal S}$ and ${\cal F}$ can be identified with the constituents of the Higgs boson, and
that their Yukawa couplings are the source of the SM Yukawa couplings, giving rise to
a flavour structure  similar to the SM structure.
Then, the one loop Feynman diagrams of  figure\fig{FeynRK}  are dressed by the underlying composite dynamic.

\subsection{Fundamental composite Higgs}\label{FPC}
Models in which the Higgs is a composite state are  prime candidates as potential source of new physics in the flavour sector~\cite{Kaplan:1983fs,Kaplan:1983sm,Dugan:1984hq}.
Fundamental theories with a Higgs as a composite state that are also able to generate SM fermion masses appeared in ref.~\cite{Sannino:2016sfx}. These theories feature both techni-scalars  ${\cal S}$ and techni-fermions ${\cal F}$.\footnote{Composite theories including TC scalars attempting to give masses to the SM quarks appeared earlier in the literature~\cite{Kagan:1991ng,Dobrescu:1995gz,Kagan:1994qg,Antola:2010nt,Antola:2011bx,Altmannshofer:2015esa}  for (walking) TC theories that didn't feature a light Higgs.}
In models of fundamental composite Higgs:
i) it is possible to replace the standard model Higgs and Yukawa sectors with a composite Higgs made of techni-particles;
ii) the SM fermion masses are generated via a partial compositeness mechanism~\cite{Kaplan:1991dc} in which the relevant composite techni-baryons emerge as bound states of a techni-fermion and a techni-scalar.

The composite theory does not address the SM naturalness issue
and it is fundamental in the sense that it can be extrapolated till the Planck scale \cite{Sannino:2016sfx}. Having a fundamental theory of composite Higgs, we use it to investigate the flavour anomalies.

\begin{table}[t!]
$$\begin{array}{|ccc|cccc|}\hline
\hbox{name} &\hbox{spin}  & \hbox{generations} & \SU(3)_c & \SU(2)_L & \U(1)_Y &\SU(3)_{\rm TC}  \\
\hline
\F_{L} &1/2&  1&1& 2 & Y=-1/2 &3  \\
\F_N^c &1/2& 1& 1 & 1 & -Y-1/2=0  &3 \\
\F_{E^c} & 1/2 &1& 1 &1 & Y-1/2=-1 &  3\\
\S_{E^c} &0& 3& 1& 1 & Y-1/2=-1 & 3   \\
\S_{D^c}& 0 &3 &  3 &1 & Y+1/6=-1/3 &  3\\
\hline
\end{array}$$
\caption{\label{tab:5frags2}\em Field content of the simplest Fundamental Composite Higgs model.
Extra fermions $\F_N^c,\F_{L}, \F_{E^c}$  with conjugated gauge quantum numbers such that the fermion content is vectorial are implicit.
Names are appropriate assuming the value $Y=-1/2$ for the  hypercharge $Y$ of ${\cal F}_L$;
however generic values are allowed.}
\label{tab:most-minimal-model}
\end{table}

The gauge group and the field content of a simple model are summarised in table~\ref{tab:most-minimal-model}.
Here the new strong group is chosen to be $\SU(N_{\rm TC})$ with $N_{\rm TC}=3$ and we list the gauge quantum numbers of the new vectorial fermions and scalars
that can provide a composite Higgs with Yukawa couplings to all SM fermions $L,E,Q,U,D$.
Three generations of techni-scalars are introduced in order to reproduce all SM fermion masses and mixings,
while having a renormalizable theory with no Landau poles below the Planck scale.
The hypercharge  $Y$ of the ${\cal F}_L$ fermion is free.
We assume the minimal choices  $Y=-1/2$,  $Y=1/2$ and $Y=0$.

The matrices of SM Yukawa couplings $y_\ell$, $y_{\rm u}$, $y_{\rm d}$ are obtained from the TC-Yukawa couplings
\beq  \label{eq:Yuk2}
\Lag_Y =
y_L \, L   \F_{L}  \S_{E^c}^*+
y_E \, E \F_{N}^c \S_{E^c} +
(y_D \, D \F_{N}^c +
y_U \, U \F_{E^c}^c) \S_{D^c}+
y_Q \, Q\F_{L}\S_{D^c}^*+
\hbox{h.c.}
\eeq
as $y_\ell \approx y_L y^T_E/g_{\rm TC}$, $y_{\rm d} \approx y_Q y_D^T/g_{\rm TC}$,
$y_{\rm u} \approx y_Q y_U^T/g_{\rm TC}$,
where the new gauge coupling $g_{\rm TC}$ becomes strong,
 $g_{\rm TC}\sim 4\pi/\sqrt{N_{\rm TC}}$, at the scale
 $\Lambda_{\rm TC} \sim g_{\rm TC} f_{\rm TC}$, forming composite particles with  mass of order $\Lambda_{\rm TC}$
  and condensates $\langle  \F\F^c\rangle \sim f_{\rm TC}^2 \Lambda_{\rm TC} $.
  In view of the resulting breaking of the TC-chiral symmetry, the Higgs doublet $H$ (identified with pseudo Goldstone bosons of the theory)  and other composite scalars remain lighter.
  Lattice simulations~\cite{Lewis:2011zb,Hietanen:2014xca,Arthur:2016dir} of the most minimal fundamental composite theories~\cite{Ryttov:2008xe,Galloway:2010bp,Cacciapaglia:2014uja}, without techni-scalars, have demonstrated the actual occurrence of chiral symmetry breaking with the relevant breaking pattern, and furthermore provided the spectrum of the spin one vector and axial techni-resonances with masses $m_V = 3.2(5)\TeV/\sin \theta$ and $m_A = 3.6(9)\TeV/\sin \theta$ where
$\theta$ is the electroweak embedding angle to be determined by the dynamics, that must be smaller than about $0.2$.

The TC-Yukawa couplings accidentally conserve lepton and baryon numbers (like in the SM) and TC-baryon number;
depending on the value of $Y$ the lightest TC-baryon can be a neutral DM candidate.

 We require TC-scalar masses and TC-quartics to respect flavour symmetries so that the BSM corrections to flavour observables abide the experimental bounds.
At one loop\footnote{The loop analysis, in the composite scenario, is merely a schematic way to keep track of the relevant factors stemming from the TC dynamics when writing SM four-fermion interactions.} in the TC-Yukawas one obtains the following operators involving 4 SM fermions
 \beq \label{eq:4fbox}
 \sum_{f,f'}^{L,E,Q,U,D}
\frac{(y_f^\dagger y_f)_{ij}(y_{f'}^\dagger y_{f'})_{i'j'}}{g_{\rm TC}^2 \Lambda_{\rm TC}^2}
(\bar f_i \gamma_\mu f'_{j'})(\bar {f}'_{i'} \gamma_\mu f_j)+
\frac{(y_L^\dagger y_E^*)_{ij}(y_Q y_D^T)_{i'j'}}{g_{\rm TC}^2 \Lambda_{\rm TC}^2}
(\bar L_i \gamma_\mu Q_{i'})(\bar E_j \gamma_\mu D_{j'}ÃŸ).\eeq
All SM fermions and their chiralities are involved.
These operators are phenomenologically viable if the fundamental TC-Yukawa couplings have the minimal values
needed to reproduce the SM Yukawa couplings:
$y_E \sim y_L \sim \sqrt{g_{\rm TC} y_\ell}$,
and similarly for quarks.

However, when the TC-Yukawas (say, $y_L$) are enhanced the impact on new physics is also enhanced. The observed SM Yukawa couplings are reproduced  when the corresponding TC-Yukawas (say, $y_E$)  are reduced.
Consequently, in this scenario new physics manifests prevalently in leptons of one given chirality. Because data prefer new physics to emerge prevalently in left-handed muons it is natural to consider here an enhanced muon coupling $y_L$ and a correspondingly reduced right-handed $y_E$.

\begin{table}
$$ \begin{array}{c|cc}
\hbox{Coefficient} & \hbox{One-loop result} & \hbox{Non-perturbative estimate}\\ \hline
c_{b_L \mu_L}
&  \displaystyle N_{\rm TC}  \frac{(y_{L} y_L^\dagger)_{\mu\mu} (y_{Q} y_{Q}^\dagger)_{bs}}{(4\pi)^2 4 M_{{\cal F}_L}^2} F(x,y)
&  \displaystyle \frac{(y_{L} y_L^\dagger)_{\mu\mu} (y_{Q} y_{Q}^\dagger)_{bs}}{g_{\rm TC}^2 \Lambda_{\rm TC}^2}\\[5mm]
c_{b_L b_L}
&  \displaystyle N_{\rm TC}  \frac{ (y_{Q} y_{Q}^\dagger)_{bs}^2}{(4\pi)^2 8 M_{{\cal F}_L}^2} F(x,x)
& \displaystyle  \frac{ (y_{Q} y_{Q}^\dagger)_{bs}^2}{g_{\rm TC}^2 \Lambda_{\rm TC}^2}\\[3mm]
\Delta a_\mu
& \displaystyle  N_{\rm TC} \frac{m_\mu^2 (y_{L} y_L^\dagger)_{\mu\mu}}{(4\pi)^2 M_{{\cal F}_L}^2 }\left[(2Y-1) F_7(y) +2Y \frac{ F_7(1/y)}{y}\right]
& \displaystyle   \frac{m_\mu^2}{{g_{\rm TC}} \Lambda_{\rm TC}^2 }\\[4mm]
\delta g_{Z \mu_L} & \displaystyle   N_{\rm TC} g_2\frac{ M_Z^2 (y_{L} y_L^\dagger)_{\mu\mu}}{2(4\pi)^2(1-2s^2_{\rm W}) M_{{\cal F}_L}^2 }{\cal F}_9 (Y,y)  & \displaystyle  g_2\frac{M_Z^2 (y_{L} y_L^\dagger)_{\mu\mu} }{g_{\rm TC}^2 \Lambda_{\rm TC}^2 }\\
  \end{array}
$$
\caption{\em\label{tab:cCH} Coefficients of the low-energy operators generated within a naive perturbative TC-fermion and TC-scalar estimate (second column) along with their NDA analysis counterpart (third column). The NDA result for $\Delta a$ modifies in the presence of a TC-fermion condensate to  $\sfrac{m_\mu v (y_{L} y_E^T)_{\mu\mu}}{{g_{\rm TC}} \Lambda_{\rm TC}^2 }$. }
\end{table}

\begin{figure}[t]
\begin{center}
\includegraphics[width=0.5\textwidth]{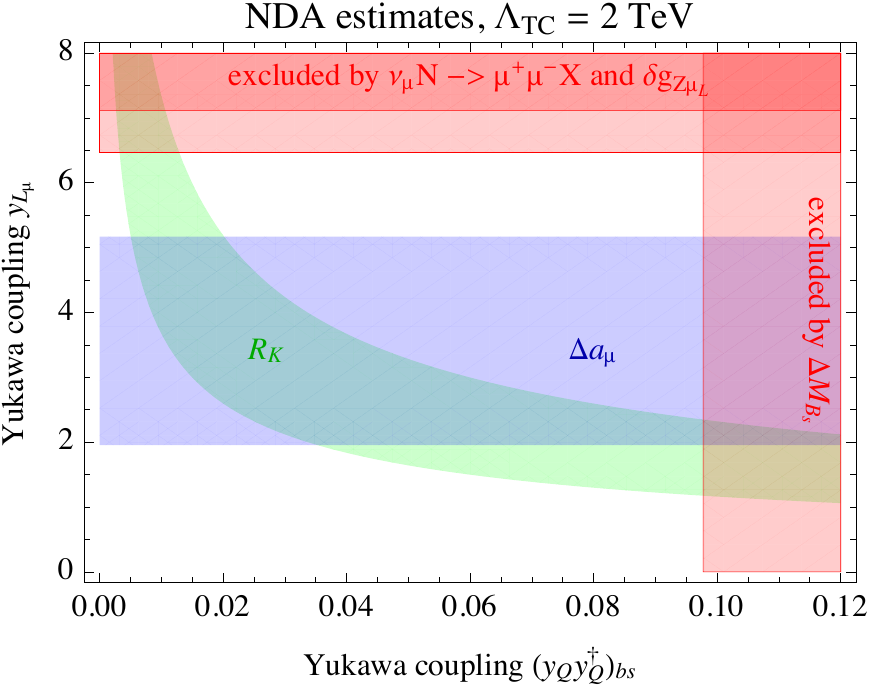}
\caption{\em Estimates of signals and bounds on the Yukawa couplings of fundamental composite Higgs models.
{The model generates an effective operator that can simultaneously account for both $R_K$ and $R_{K^*}$, so
only $R_K$ is plotted.}
\label{fig:CompositeH}}
\end{center}
\end{figure}
We summarise in table~\ref{tab:cCH} the   coefficients of the relevant flavour-violating effective operators,
both within a naive one-loop approximation
(adopting the results from ref.~\cite{1408.4097,1608.07832})
and Naive Dimensional Analysis (NDA) in the composite theory.
We defined  $x=\sfrac{M_{{\cal S}_D^c}^2}{M_{{\cal F}_L}^2}$, $y =\sfrac{M_{{\cal S}_E^c}^2}{M_{{\cal F}_L}^2}$
and the loop functions
\begin{eqnsystem}{sys:loopf}
F(x,y) &=& \frac{1}{(1-x)(1-y)} + \frac{x^2 \ln x}{(1-x)^2(x-y)} + \frac{y^2 \ln y}{(1-y)^2(y-x)}\\[3mm]
F_7(y) &=& \frac{y^3-6y^2+6y\ln y+3y+2}{12(1-y)^4},\\
{F}_9 (y) &=&  \frac{-2y^3 + 6 \ln y + 9 y^2 - 18y + 11}{36(y-1)^4}\ , \\
{G}_9 (y) &=&  \frac{7  -36 y + 45 y^2 - 16y^3 + 6(2y-3)y^2\ln y}{36(y-1)^4},\\
{\cal F}_9 (Y,y) &=&  s^2_{\rm W}(2Y - 1) F_9(y) - (1 -s^2_{\rm W}(2Y+1))G_9(y)
\end{eqnsystem}
that equal $F(1,1) = 1/3$, $F_7(1)=\tilde F_7(1)=1/24$,
${F}_9 (1) = -\sfrac{1}{24}$, $G_9 (1) = \sfrac{1}{8} $,
for degenerate masses.
The latter entry in table~\ref{tab:cCH}  is the correction to the $Z$ coupling to left-handed muons $g_{\mu_L}$, written in terms of the   weak mixing angle
$s_{\rm W} = \sin \theta_{\rm W}$.
The LEP bound at the $Z$ pole is $|\delta g_{Z \mu_L}| \leq 0.8\% \cdot  g_2$ at $2\sigma$~\cite{hep-ex/0509008}.
We can neglect TC-penguin diagrams~\cite{1408.4097}.
We can always work in a basis where $y_L=\diag(y_{L_e}, y_{L_\mu}, y_{L_\tau})$ is diagonal, such that
$(y_{L} y_L^\dagger)_{\mu\mu} = y_{L_\mu}^2$.

\medskip

Figure\fig{CompositeH} shows that,
in order to reproduce the $b\to s\ell^+\ell^-$ anomalies and the muon $g-2$ anomaly,
a relatively large Yukawa coupling $y_{L_\mu}\sim 1.5$ is needed, like in models with perturbative extra fermions and scalars.
In the composite model such values of TC-Yukawa coupling have natural sizes.
This is corroborated by a RGE analysis for $y_{L_\mu}$ that features an extra contribution involving the $g_{\rm TC}$ gauge coupling:
\beq  (4\pi)^2 \frac{\partial y_{L_\mu}}{\partial\ln\mu}=\frac{N_{\rm TC} +3 }{2} y_{L_\mu}^3 -3 \frac{N_{\rm TC}^2-1}{2N_{\rm TC}} g_{\rm TC}^2y_{L_\mu} \, ,
\eeq
In the presence of the first term only, setting $N_{\rm TC}=1$,
the Yukawa coupling grows with energy. Perturbativity up to a scale $\Lambda_{\rm max}$
implies
$|y_{L_\mu}| < 2\pi/\sqrt{\ln(\Lambda_{\rm max}/\TeV)}$,
with $|y_{L_\mu}| \approx 1$ for $\Lambda_{\rm max} \sim M_{\rm Pl}$.
In the presence of the second term a larger $y_{L_\mu}\sim g_{\rm TC}$ is compatible with the requirement that
all couplings can be extrapolated up to the Planck scale.
This is similar to how the strong coupling $g_3$ allows for $y_t \approx 1$ in the SM.
In the fundamental composite Higgs model, the large couplings $y_t$ and $y_{L_\mu}$
contribute to the prediction for the Higgs mass parameter in terms
of $\Lambda_{\rm TC}$.

\medskip

Lepton-flavour violation is absent as long as the $y_E$ matrix is diagonal in the same basis where $y_L$ is diagonal.
Then $y_\ell= y_{L_\ell} y_{E_\ell}/g_{\rm TC}$ for $\ell=\{e,\mu,\tau\}$.
In general, there can be a flavour-violating mixing matrix in the lepton sector.
In particular, the mixing angle $\theta_{e\mu}$ generates $\mu\to e\gamma$, but only
when effects at higher order in the Yukawa couplings are included~\cite{Sannino:2016sfx}.
Focusing on effects enhanced by the large coupling $y_{L_\mu}$ one has
\beq \hbox{BR}(\mu\to e\gamma) \sim
\frac{4\pi \alpha_{\rm em} v^6 y_{E_e}^2 y_{L_\mu}^6\theta_{e\mu}^2}{g_{\rm TC}^6 m_\mu^2 \Lambda_{\rm TC}^4}
\sim
y_{E_e}^2 y_{L_\mu}^6 \theta_{e\mu}^2 \left(\frac{2\TeV}{\Lambda_{\rm TC}}\right)^4\eeq
The experimental bound $\hbox{BR}(\mu\to e\gamma) < 0.6 ~10^{-12}$~\cite{MEG} is satisfied even for $\theta_{e\mu}\sim 1$
provided that in the electron sector too one has a large $y_{L_e}$ and a small $y_{E_e} \sim y_e \sim 10^{-6}$.

Finally, we mention an effect that can enhance the new-physics correction to some flavour-violating operators.
While the fermion condensates induced by the strong dynamics are known,
the scalar condensates are not known (although perhaps they are computable, for example by dedicated lattice simulations).
Possible scalar condensates could break the accidental flavour symmetry among scalars, leading to extra lighter
composite pseudo-Goldstone bosons.  The state made of ${\cal S}_{E^c}^* {\cal S}_{D^c}$ behaves as a lepto-quark:
if light it would mediate at tree level  some effective operators, analogously to the $\tilde{S}_1$ lepto-quark considered in section~\ref{LQ}.

\section{Conclusions}\label{concl}
We found that the new measurement of $R_{K^*}$ together with $R_K$ favours new physics in left-handed leptons.
Furthermore, adding to the fit kinematical $b\to s \mu^+\mu^-$ distributions
(affected by theoretical uncertainties), one finds that they favour similar deviations from the SM in left-handed muons.
However, even if the {experimental} uncertainties on $R_K$, $R_{K^*}$ will be reduced,
a precise determination of the new-physics parameters will be prevented by the fact that these are no longer theoretically clean observables, if new physics really affects muons differently from electrons.

We next discussed possible theoretical interpretations of the anomaly.
One can build models compatible with all other data:
\begin{itemize}
\item One extra $Z'$ vector can give extra new-physics operators that involve all chiralities of SM leptons.
The simplest possibility motivated by anomaly cancellation  is a vectorial coupling to leptons.
However, unless the $Z'$ is savagely coupled to $\bar b s$ quarks, a $Z'$ coupled to $\bar s s$ and $\bar b b$
is disfavoured by $pp \to Z'\to \mu^+\mu^-$ searches at LHC and other contraints.

\item
One lepto-quark tends to give effects in muons or electron only (in order to avoid large flavour violations), and only in one chirality.
\item One can add extra fermions and scalars such that they mediate, at one loop level, the desired new physics.
Their Yukawa coupling to muons must be larger than unity.
\end{itemize}

\smallskip

While the effective 4-fermion operators that can account for the $b\to s\ell^+\ell^-$ anomalies need to be suppressed by a scale $\sim 30\TeV$,
the actual new physics can be at a lower scale, with obvious consequences for direct observability at the LHC and for Higgs mass naturalness.

\small

\subsubsection*{Acknowledgments}
We thank Marina Marinkovic, Andrea Tesi and Elena Vigiani for useful discussions. We thank Javier Virto for further useful comments.
The work of A.S.\ is supported by  the ERC grant NEO-NAT, the one of F.S.\ is partially supported by the Danish National Research Foundation Grant DNRF:90, and the one of
G. D'A. thanks Andrea Wulzer and the Lattice QCD group of CERN, and in particular Agostino Patella, for sharing their computing resources.
 R.T.\ is supported by the Swiss National Science Foundation under the grant CRSII2-160814 (Sinergia).
R.T.\ thanks INFN Sezione di Genova for computing resources, Alessandro Brunengo for computing support, and David Straub for clarifications about the {\sc Flavio} program.

\appendix
\section{List of observables used in the global fit}\label{appendixA}
In table~\ref{tab:ang_obs} and \ref{tab:br_obs} we summarize the observables used in addition to the `clean' observables. All bins are treated in the experimental analyses as independent, even if overlapping. It is clear that a correlation should exists between measurements in overlapping bins, however this is not estimated by the experimental collaborations. For this reason we include in our fit the measurements in all relevant bins, even if overlapping, without including any correlation beyond the ones given in the experimental papers. Notice that, for instance in the case of the LHCb analysis \cite{1512.04442}, the result in the bin $[1.1,6]$ GeV$^2$ has a smaller error than the measurements in the bins $[1.1,2.5],[2.5,4],[4,6]$ GeV$^2$, even when the information from these three bins is combined. In fact, we verified that the bin $[1.1,6]$ GeV$^2$ has a stronger impact on our fits than the three smaller bins. This shows that even if the measurements are potentially largely correlated, the largest bin dominates the fit, so that the effect of the unknown correlation becomes negligible.

\begin{table}[!tb!]
\begin{center}
\small\begin{tabular}{||c|c||}
\hline\hline
  \multicolumn{2}{|c|}{\textbf{Angular observables}}   \\
\hline\hline
\cline{1-2}{{Observable}}  & {{$[q^{2}_{\text{min}},q^{2}_{\text{max}}]$ [GeV$^{2}$]}} \\ \hline
\multicolumn{2}{||c||}{{\color{black}{LHCb $B\to K^{*}\mu\mu$ 2015 S \cite{1512.04442}}}} \\
\hline
\text{$\langle F_{L} \rangle (B^{0}\to K^{*}\mu\mu)$} & [1.1, 6], [15, 19], [0.1, 0.98], [1.1, 2.5], [2.5, 4], [4, 6], [15, 17], [17, 19] \\  \hline
 \text{$\langle S_{3} \rangle (B^{0}\to K^{*}\mu\mu)$} & [1.1, 6], [15, 19], [0.1, 0.98], [1.1, 2.5], [2.5, 4], [4, 6], [15, 17], [17, 19] \\  \hline
  \text{$\langle S_{4} \rangle (B^{0}\to K^{*}\mu\mu)$} & [1.1, 6], [15, 19], [0.1, 0.98], [1.1, 2.5], [2.5, 4], [4, 6], [15, 17], [17, 19] \\  \hline
 \text{$\langle S_{5} \rangle (B^{0}\to K^{*}\mu\mu)$} & [1.1, 6], [15, 19], [0.1, 0.98], [1.1, 2.5], [2.5, 4], [4, 6], [15, 17], [17, 19] \\  \hline
 \text{$\langle S_{7} \rangle (B^{0}\to K^{*}\mu\mu)$} & [1.1, 6], [15, 19], [0.1, 0.98], [1.1, 2.5], [2.5, 4], [4, 6], [15, 17], [17, 19] \\  \hline
 \text{$\langle S_{8} \rangle (B^{0}\to K^{*}\mu\mu)$} & [1.1, 6], [15, 19], [0.1, 0.98], [1.1, 2.5], [2.5, 4], [4, 6], [15, 17], [17, 19] \\  \hline
  \text{$\langle S_{9} \rangle (B^{0}\to K^{*}\mu\mu)$} & [1.1, 6], [15, 19], [0.1, 0.98], [1.1, 2.5], [2.5, 4], [4, 6], [15, 17], [17, 19] \\  \hline
  \text{$\langle  A_{FB} \rangle (B^{0}\to K^{*}\mu\mu)$} & [1.1, 6], [15, 19], [0.1, 0.98], [1.1, 2.5], [2.5, 4], [4, 6], [15, 17], [17, 19] \\  \hline \hline
  \multicolumn{2}{||c||}{{\color{black}{CMS $B\to K^{*}\mu\mu$ 2017 \cite{CMS:2017ivg}}}} \\
\hline
\text{$\langle P_{1} \rangle (B^{0}\to K^{*}\mu\mu)$} & [1, 2], [2, 4.3], [4.3, 6], [16, 19] \\  \hline
 \text{$\langle P_{5}' \rangle (B^{0}\to K^{*}\mu\mu)$} & [1, 2], [2, 4.3], [4.3, 6], [16, 19] \\  \hline \hline
    \multicolumn{2}{||c||}{{\color{black}{ATLAS $B\to K^{*}\mu\mu$ 2017 \cite{ATLAS:2017dlm}}}} \\
\hline
\text{$\langle F_{L} \rangle (B^{0}\to K^{*}\mu\mu)$} & [0.04, 2], [2, 4], [4, 6], [0.04, 4], [1.1, 6], [0.04, 6] \\  \hline
 \text{$\langle S_{3} \rangle (B^{0}\to K^{*}\mu\mu)$} & [0.04, 2], [2, 4], [4, 6], [0.04, 4], [1.1, 6], [0.04, 6] \\  \hline
 \text{$\langle S_{4} \rangle (B^{0}\to K^{*}\mu\mu)$} & [0.04, 2], [2, 4], [4, 6], [0.04, 4], [1.1, 6], [0.04, 6] \\  \hline
 \text{$\langle S_{5} \rangle (B^{0}\to K^{*}\mu\mu)$} & [0.04, 2], [2, 4], [4, 6], [0.04, 4], [1.1, 6], [0.04, 6] \\  \hline
 \text{$\langle S_{7} \rangle (B^{0}\to K^{*}\mu\mu)$} & [0.04, 2], [2, 4], [4, 6], [0.04, 4], [1.1, 6], [0.04, 6] \\  \hline
 \text{$\langle S_{8} \rangle (B^{0}\to K^{*}\mu\mu)$} & [0.04, 2], [2, 4], [4, 6], [0.04, 4], [1.1, 6], [0.04, 6] \\  \hline
 \text{$\langle P_{1} \rangle (B^{0}\to K^{*}\mu\mu)$} & [0.04, 2], [2, 4], [4, 6], [0.04, 4], [1.1, 6], [0.04, 6] \\  \hline
 \text{$\langle P_{4}' \rangle (B^{0}\to K^{*}\mu\mu)$} & [0.04, 2], [2, 4], [4, 6], [0.04, 4], [1.1, 6], [0.04, 6] \\  \hline
 \text{$\langle P_{5}' \rangle (B^{0}\to K^{*}\mu\mu)$} & [0.04, 2], [2, 4], [4, 6], [0.04, 4], [1.1, 6], [0.04, 6] \\  \hline
 \text{$\langle P_{6}' \rangle (B^{0}\to K^{*}\mu\mu)$} & [0.04, 2], [2, 4], [4, 6], [0.04, 4], [1.1, 6], [0.04, 6] \\  \hline
 \text{$\langle P_{8}' \rangle (B^{0}\to K^{*}\mu\mu)$} & [0.04, 2], [2, 4], [4, 6], [0.04, 4], [1.1, 6], [0.04, 6] \\  \hline\hline
\hline

   \end{tabular}
\end{center}
\caption{\em List of angular observables used in the global fit in addition to the `clean' observables.
\label{tab:ang_obs}}
\end{table}%
\normalsize

\begin{table}[!htb!]
\begin{center}
\small\begin{tabular}{||c|c||}
\hline\hline
  \multicolumn{2}{|c|}{\textbf{Branching ratios}}   \\
\hline\hline
\cline{1-2}{{Observable}}  & {{$[q^{2}_{\text{min}},q^{2}_{\text{max}}]$ [GeV$^{2}$]}} \\ \hline

\multicolumn{2}{||c||}{{\color{black}{LHCb $B^{\pm}\to K\mu\mu$ 2014 \cite{1403.8044}}}} \\
\hline
\multirow{2}{*}{$\frac{d}{dq^{2}}\text{BR}(B^{\pm}\to K\mu\mu)$} & [0.1, 0.98], [1.1, 2], [2, 3], [3, 4], [4, 5], [5, 6], [15, 16], [16, 17],\\
& [17, 18], [18, 19], [19, 20], [20, 21], [21, 22], [1.1, 6], [15, 22]  \\  \hline \hline

\multicolumn{2}{||c||}{{\color{black}{LHCb $B^{0}\to K\mu\mu$ 2014 \cite{1403.8044}}}} \\
\hline
$\frac{d}{dq^{2}}\text{BR}(B^{0}\to K\mu\mu)$ & [0.1, 2], [2, 4], [4, 6], [15, 17], [17, 22], [1.1, 6], [15, 22]\\  \hline \hline

\multicolumn{2}{||c||}{{\color{black}{LHCb $B^{\pm}\to Kee$ 2014 \cite{1406.6482}}}} \\
\hline
$\frac{d}{dq^{2}}\text{BR}(B^{\pm}\to K ee)$ & [1, 6]\\  \hline \hline

\multicolumn{2}{||c||}{{\color{black}{LHCb $B^{\pm}\to K^{*}\mu\mu$ 2014 \cite{1403.8044}}}} \\

\hline
$\frac{d}{dq^{2}}\text{BR}(B^{\pm}\to K^{*}\mu\mu)$ & [0.1, 2], [2, 4], [4, 6], [15, 17], [17, 19], [1.1, 6], [15, 19] \\  \hline \hline

\multicolumn{2}{||c||}{{\color{black}{LHCb $B^{0}\to K^{*}\mu\mu$ 2016 \cite{1606.04731}}}} \\
\hline
$\frac{d}{dq^{2}}\text{BR}(B^{0}\to K^{*}\mu\mu)$ & [0.1, 0.98], [1.1, 2.5], [2.5, 4], [4, 6], [15, 17], [17, 19], [1.1, 6], [15, 19] \\  \hline \hline

\multicolumn{2}{||c||}{{\color{black}{LHCb $B_{s}\to \phi\mu\mu$ 2015 \cite{1506.08777}}}} \\
\hline
$\frac{d}{dq^{2}}\text{BR}(B_{s}\to \phi \mu\mu)$ & [0.1, 2], [2, 5], [15, 17], [17, 19], [1, 6], [15, 19] \\  \hline \hline

\multicolumn{2}{||c||}{{\color{black}{BaBar $B \to X_{s}ll$ 2013 \cite{1312.5364}}}} \\
\hline
$\frac{d}{dq^{2}}\text{BR}(B\to X_{s} \mu\mu)$ & [1, 6], [0.1, 2], [2, 4.3], [4.3, 6.8], [14.2, 25] \\  \hline
$\frac{d}{dq^{2}}\text{BR}(B\to X_{s} ee)$ & [1, 6], [0.1, 2], [2, 4.3], [4.3, 6.8], [14.2, 25] \\  \hline
$\frac{d}{dq^{2}}\text{BR}(B\to X_{s} ll)$ & [1, 6], [0.1, 2], [2, 4.3], [4.3, 6.8], [14.2, 25] \\  \hline \hline

\multicolumn{2}{||c||}{{\color{black}{Belle $B \to X_{s}ll$ 2005 \cite{hep-ex/0503044}}}} \\
\hline
$\frac{d}{dq^{2}}\text{BR}(B\to X_{s} ll)$ & [0.04, 1], [1, 6], [14.4, 25]\\  \hline \hline

\hline

   \end{tabular}
\end{center}
\caption{\em List of differential branching ratios used in the global fit in addition to the `clean' observables.
\label{tab:br_obs}}
\end{table}%
\normalsize

\footnotesize\begin{multicols}{2}

\end{multicols}

\newpage

\normalsize

~

\begin{figure}[t]
\minipage{0.5\textwidth}
  \includegraphics[width=0.95\textwidth]{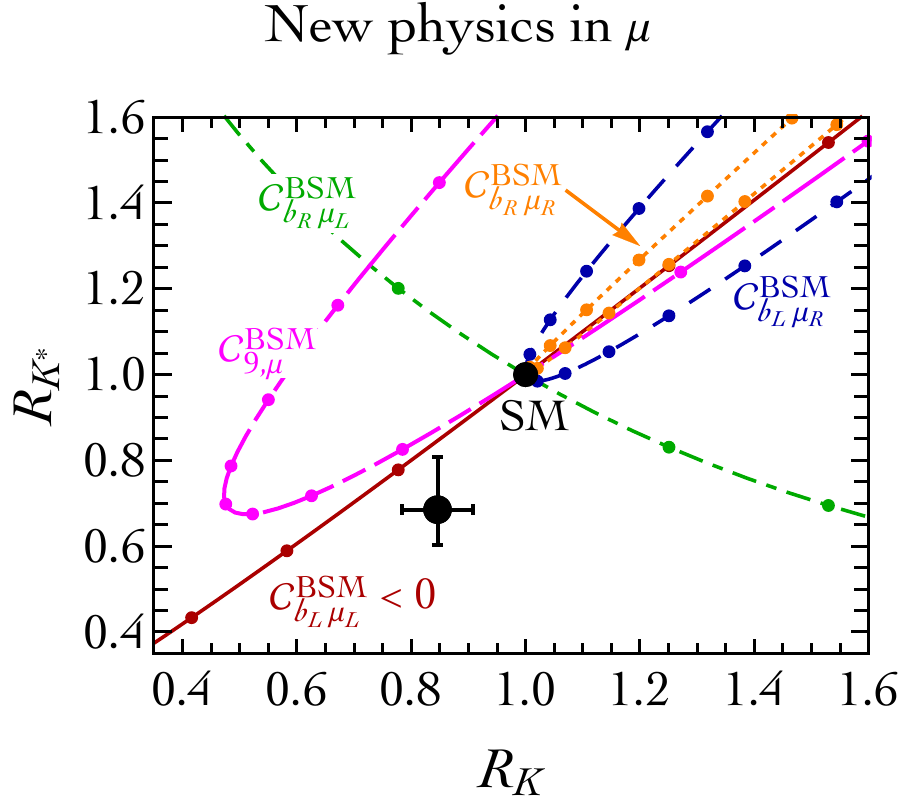}
\endminipage 
\minipage{0.5\textwidth}
  \includegraphics[width=0.95\textwidth]{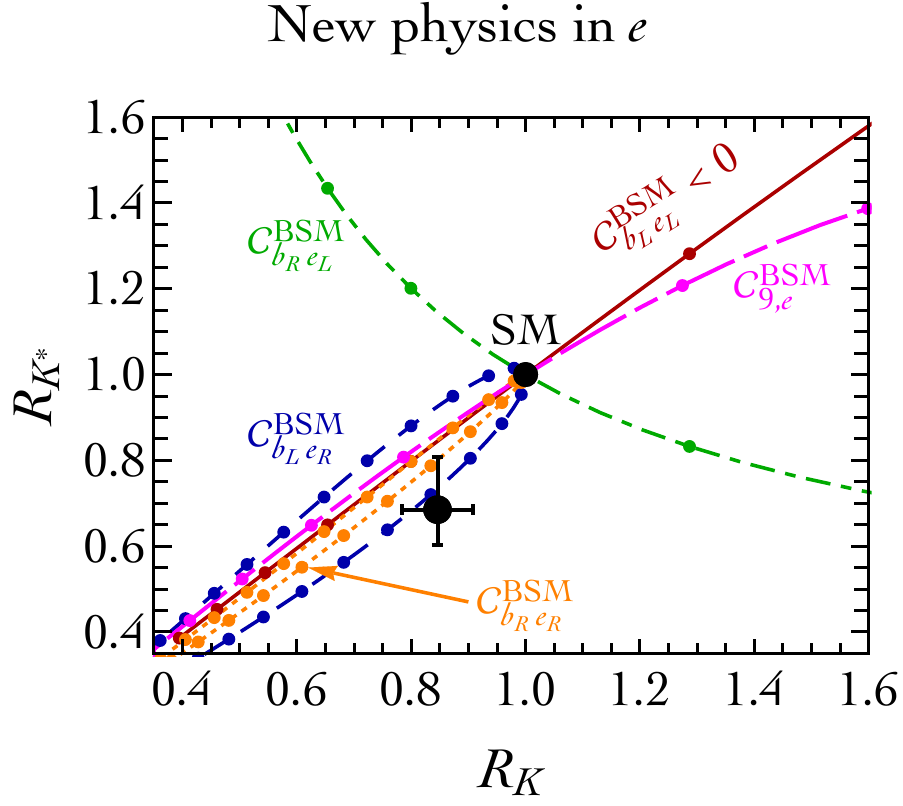}
\endminipage
\caption{\em As in fig.~\ref{fig:RKRKStar}, adding Moriond EW 2019  data about $R_K$
and $R_{K^*}$. 
Black: combination. 
Small dots on the theoretical predictions mark changes in steps of $1$ of the corresponding 
coefficients.
\label{fig:RKRKStarUpdate}}
\end{figure}

\section*{6~~~Addendum: results presented at Moriond EW 2019}\label{6in}
The LHCb collaboration presented a new measurement of
the $R_K$ ratio using a part of LHC run II data, and a revised analysis of run I data.
The result~\cite{LHCBMor19}
\beq R_K|_{\rm run~I}^{\rm revised} = 0.717^{+0.083}_{-0.071} \pm 0.017,\qquad
R_K|_{\rm run~II} = 0.928^{+0.089}_{-0.076}\pm0.020,\eeq
is compatible both with the SM and with the previous result.
The average is~\cite{LHCBMor19}
\beq \label{eq:RKexp}
R_K =\frac{{\rm BR} \left ( B^+ \to K^+ \mu ^+ \mu ^-\right )}{{\rm BR} \left ( B^+ \to K^+ e^+ e^-\right )}= 
0.846^{+0.060}_{-0.054} \pm0.015
\eeq
The central value is $2.5$ standard deviations below unity.
Furthermore, the {\sc Belle} collaboration presented preliminary 
results about $R_{K^*}$.  Averaging over $B^0$ and $B^+$, 
the {\sc Belle} result for the quantity defined in eq.\eq{RKq} is~\cite{belleMor19}.
\beq R_{K^*}[0.045,1.1]=0.52^{+0.36}_{-0.26}\pm0.05,\qquad
R_{K^*}[1.1,6]=0.96^{+0.45}_{-0.29}\pm0.11.\eeq
We combine these data with the LHCb data in eq.\eq{RK*LHCB},
which have smaller uncertainties.
At the light of these new data
we update our previous results concerning both 
the statistical significance of the anomaly as well
the preferred chiral structure of the possible new physics.
With the new data, the statistical significance of the anomaly remains
essentially unchanged.

\begin{figure}[!htb!]
\begin{center}
\includegraphics[width=0.99\textwidth]{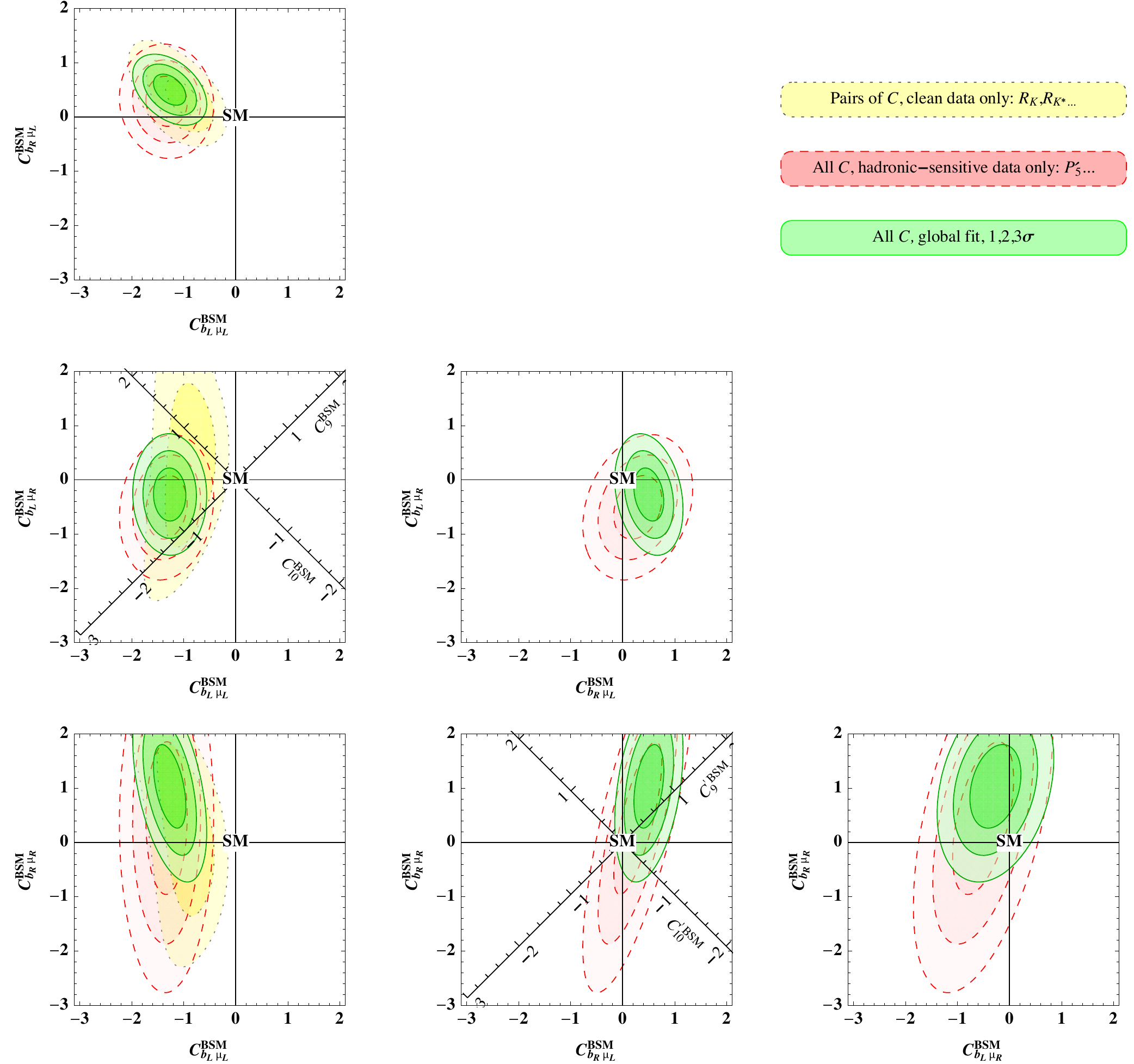}
\caption{\em As in fig.~\ref{fig:fitglobalmu}, adding Moriond EW 2019 data about $R_K$ and $R_{K^*}$.
\label{fig:fitglobalmu19}}
\end{center}
\end{figure}

\begin{figure}[tp]
\begin{center}
\includegraphics[width=0.99\textwidth]{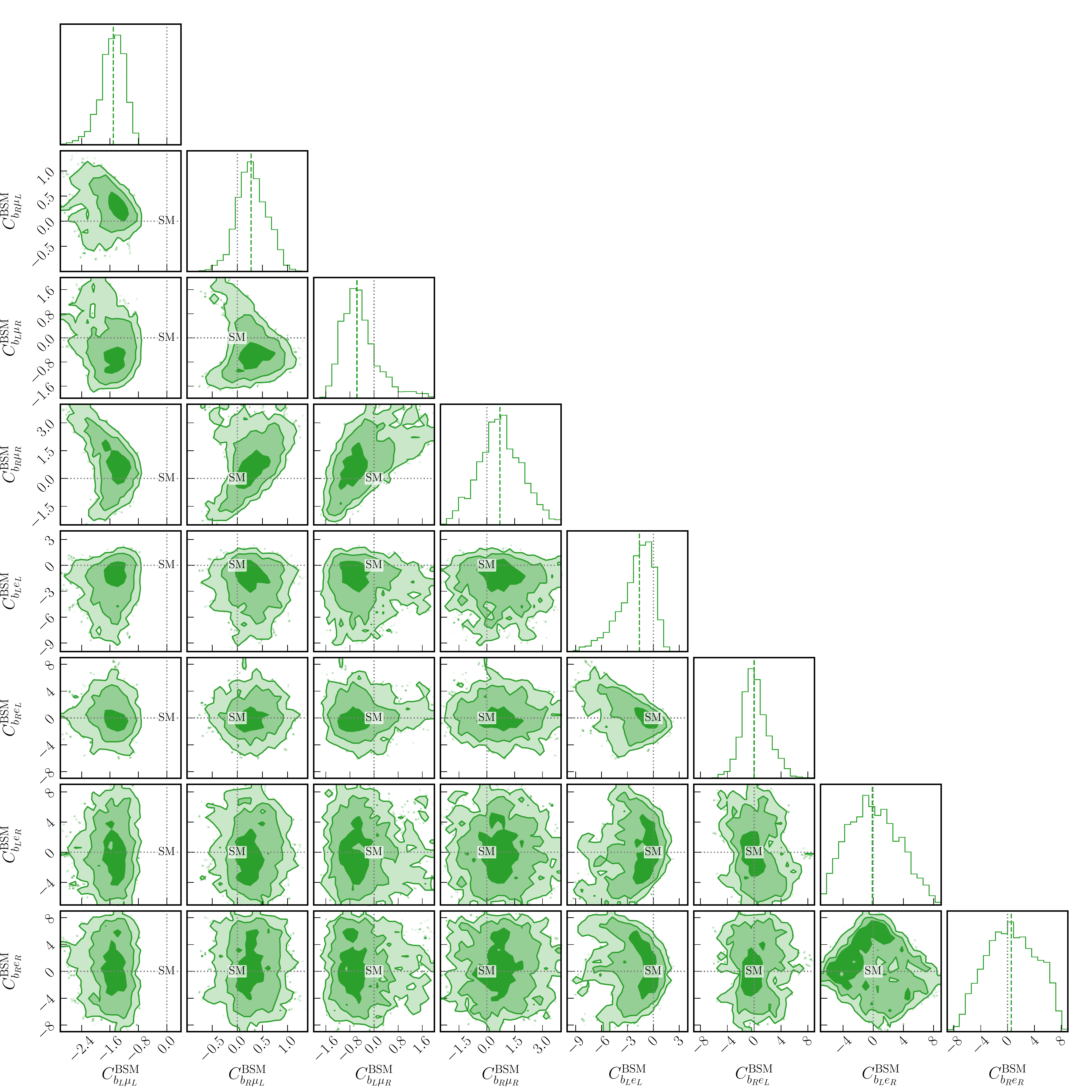}
\caption{\em As in fig.~\ref{fig:fitglobalmue}, adding Moriond EW 2019 data about $R_K$ and $R_{K^*}$.
\label{fig:fitglobalmue2019}}
\end{center}
\end{figure}

Assuming new physics in muons only,
fig.\fig{fitglobalmu} becomes fig.\fig{fitglobalmu19} and
the best fits of eq.\eq{muold} become
\beq
\begin{array}{rcl}
C_ {b_L\mu _L}^{\text{BSM}} &=& -1.27 \pm 0.21\\ 
C_ {b_R\mu _L}^{\text{BSM}} &=& +0.50 \pm 0.19\\ 
C_ {b_L\mu _R}^{\text{BSM}} &=& -0.27 \pm 0.32\\ 
C_ {b_R\mu _R}^{\text{BSM}} &=& +1.04 \pm 0.51\\ 
\end{array}\qquad\hbox{with}\qquad\rho = 
\left(
\begin{array}{cccc}
 1 & -0.41 & -0.02 & -0.48 \\
 -0.41 & 1 & -0.25 & 0.38 \\
 -0.02 & -0.25 & 1 & 0.29 \\
 -0.48 & 0.38 & 0.29 & 1 \\
\end{array}
\right)
.\eeq
Assuming new physics in muons and electrons, 
fig.\fig{fitglobalmue} becomes fig.\fig{fitglobalmue2019},
table~\ref{tab:FARTFit} and table~\ref{tab:FARTFit2} become table~\ref{tab:FARTFit2Update}.
See~\cite{1903.09578,1903.09617,1903.09632} for dedicated fits.

\begin{table}[pt]
\begin{center}
\begin{tabular}{||c||c|c|c||c|c|c||c|c|c||}
\hline\hline
  \multicolumn{10}{|c|}{\textbf{New physics in the muon sector}}   \\
\hline\hline
{\color{blue}{Wilson}} & \multicolumn{3}{c||}{{\color{blue}{Best-fit}}} & \multicolumn{3}{c||}{{\color{blue}{1-$\sigma$ range}}} &
 \multicolumn{3}{c||}{{\color{blue}{$\sqrt{\chi^2_{\rm SM} - \chi^2_{\rm best}}$}}}   \\ \cline{2-10}
{\color{blue}{coeff.}} &  {\color{giallo}{`clean'}}  & {\color{rossoc}{`HS'}} & {\color{verde}{all}}  & {\color{giallo}{`clean'}}  & {\color{rossoc}{`HS'}} & {\color{verde}{all}} & {\color{giallo}{`clean'}} & {\color{rossoc}{`HS'}} & {\color{verde}{all}}    \\ \hline
  \multirow{2}{*}{$C_{b_L \mu_L}^{\rm BSM}$} &  
  \multirow{2}{*}{$-0.87$} & \multirow{2}{*}{$-1.33$} &  \multirow{2}{*}{$-1.05$}
    & $-0.63$ & $-1.01$  & $-0.86$ &
     \multirow{2}{*}{$3.8$} &  \multirow{2}{*}{$4.6$} & \multirow{2}{*}{$5.9$}   \\
  & & & & $-1.11$ & $-1.68$ & $-1.24$ & & & \\  \hline\hline
  \multirow{2}{*}{$C_{b_L \mu_R}^{\rm BSM}$} & 
   \multirow{2}{*}{$0.56$} & \multirow{2}{*}{$-0.73$} &  \multirow{2}{*}{$-0.29$} 
    & $1.05$ & $-0.40$  & $0.03$ &
     \multirow{2}{*}{$1.1$} &  \multirow{2}{*}{$2.1$} & \multirow{2}{*}{$0.9$}   \\
  & & & & $0.07$ & $-1.03$ & $-0.58$ & & & \\  \hline\hline
    \multirow{2}{*}{$C_{b_R \mu_L}^{\rm BSM}$} &  
    \multirow{2}{*}{$-0.10$} & \multirow{2}{*}{$-0.20$} &  \multirow{2}{*}{$-0.16$} 
     & $0.12$ & $-0.04$  & $-0.04$ & 
     \multirow{2}{*}{$0.4$} &  \multirow{2}{*}{$1.3$} & \multirow{2}{*}{$1.3$}   \\
  & & & & $-0.31$ & $-0.29$ & $-0.26$ & & & \\  \hline\hline
      \multirow{2}{*}{$C_{b_R \mu_R}^{\rm BSM}$} & 
       \multirow{2}{*}{$-0.41$} & \multirow{2}{*}{$0.41$} &  \multirow{2}{*}{$0.26$} 
        & $0.08$ & $0.61$  & $0.47$ & 
        \multirow{2}{*}{$0.8$} &  \multirow{2}{*}{$1.7$} & \multirow{2}{*}{$1.1$}   \\
  & & & & $-0.91$ & $0.18$ & $0.04$ & & & \\  \hline\hline
   \multicolumn{10}{|c|}{\textbf{New physics in the electron sector}}   \\
\hline\hline
{\color{blue}{Wilson}} & \multicolumn{3}{c||}{{\color{blue}{Best-fit}}} & \multicolumn{3}{c||}{{\color{blue}{1-$\sigma$ range}}} &
\multicolumn{3}{c||}{{\color{blue}{$\sqrt{\chi^2_{\rm SM} - \chi^2_{\rm best}}$}}}   \\ \cline{2-10}
{\color{blue}{coeff.}} &  {\color{giallo}{`clean'}}  & {\color{rossoc}{`HS'}} & {\color{verde}{all}}  & {\color{giallo}{`clean'}}  & {\color{rossoc}{`HS'}} & {\color{verde}{all}} & {\color{giallo}{`clean'}} & {\color{rossoc}{`HS'}} & {\color{verde}{all}}    \\ \hline
     \multirow{2}{*}{$C_{b_L e_L}^{\rm BSM}$} & 
     \multirow{2}{*}{$1.01$} & \multirow{2}{*}{$0.15$} &  \multirow{2}{*}{$0.79$} 
      & $1.35$ & $0.69$  & $1.05$ & 
      \multirow{2}{*}{$3.7$} &  \multirow{2}{*}{$0.3$} & \multirow{2}{*}{$3.5$}   \\
 & & & & $0.71$ & $-0.39$ & $0.55$ & & & \\  \hline\hline
     \multirow{2}{*}{$C_{b_L e_R}^{\rm BSM}$} & 
      \multirow{2}{*}{$-4.15$} & \multirow{2}{*}{$-1.70$} &  \multirow{2}{*}{$-3.31$} 
       & $-3.45$ & $0.33$  & $-2.72$ & 
       \multirow{2}{*}{$4.1$} &  \multirow{2}{*}{$0.9$} & \multirow{2}{*}{$3.8$}   \\
 & & & & $-4.83$ & $-2.83$ & $-3.84$ & & & \\  \hline\hline
      \multirow{2}{*}{$C_{b_R e_L}^{\rm BSM}$} & 
       \multirow{2}{*}{$0.18$} & \multirow{2}{*}{$-0.51$} &  \multirow{2}{*}{$0.13$}  & 
       $0.41$ & $0.29$  & $0.35$ & 
       \multirow{2}{*}{$0.8$} &  \multirow{2}{*}{$0.7$} & \multirow{2}{*}{$0.6$}   \\
 & & & & $-0.04$ & $-1.55$ & $-0.08$ & & & \\  \hline\hline
      \multirow{2}{*}{$C_{b_R e_R}^{\rm BSM}$} &
        \multirow{2}{*}{$-4.39$} & \multirow{2}{*}{$2.10$} &  \multirow{2}{*}{$-3.32$} 
         & $-3.69$ & $3.52$  & $-2.53$ & 
         \multirow{2}{*}{$3.9$} &  \multirow{2}{*}{$0.5$} & \multirow{2}{*}{$2.7$}   \\
 & & & & $-5.08$ & $-2.69$ & $-3.99$ & & & \\  \hline\hline
  \multicolumn{10}{|c|}{\textbf{New physics in the muon sector (Vector Axial basis)}}   \\
\hline\hline
{\color{blue}{Wilson}} & \multicolumn{3}{c||}{{\color{blue}{Best-fit}}} & \multicolumn{3}{c||}{{\color{blue}{1-$\sigma$ range}}} &
 \multicolumn{3}{c||}{{\color{blue}{$\sqrt{\chi^2_{\rm SM} - \chi^2_{\rm best}}$}}}   \\ \cline{2-10}
{\color{blue}{coeff.}} &  {\color{giallo}{`clean'}}  & {\color{rossoc}{`HS'}} & {\color{verde}{all}}  & {\color{giallo}{`clean'}}  & {\color{rossoc}{`HS'}} & {\color{verde}{all}} & {\color{giallo}{`clean'}} & {\color{rossoc}{`HS'}} & {\color{verde}{all}}    \\ \hline
  \multirow{2}{*}{$C_{9,\,\mu}^{\rm BSM}$} &  
  \multirow{2}{*}{$\rm -0.89$} & \multirow{2}{*}{$-1.15$} &  \multirow{2}{*}{$-1.09$}  
  & $-1.19$ & $-0.98$  & $-0.94$ & \multirow{2}{*}{$3.5$} &  \multirow{2}{*}{$5.5$} & \multirow{2}{*}{$6.5$}   \\
  & & & & $-1.34$ & $-1.31$ & $-1.24$ & & & \\  \hline\hline
  \multirow{2}{*}{$C_{10,\,\mu}^{\rm BSM}$} &  
  \multirow{2}{*}{$0.78$} & \multirow{2}{*}{$0.48$} &  \multirow{2}{*}{$0.62$}  
  & $1.02$ & $0.69$ & $0.78$ 
  & \multirow{2}{*}{$3.8$} &  \multirow{2}{*}{$2.4$} & \multirow{2}{*}{$4.4$}   \\
  & & & & $0.56$ & $0.28$ & $0.47$ & & & \\  \hline\hline
    \multirow{2}{*}{$C_{9,\,\mu}^{\prime{\rm BSM}}$} &  
    \multirow{2}{*}{$-0.17$} & \multirow{2}{*}{$-0.24$} &  \multirow{2}{*}{$-0.22$}  
    & $0.04$ & $-0.15$  & $-0.15$ &
     \multirow{2}{*}{$0.8$} &  \multirow{2}{*}{$1.7$} & \multirow{2}{*}{$1.9$}   \\
  & & & & $-0.39$ & $-0.37$ & $-0.33$ & & & \\  \hline\hline
      \multirow{2}{*}{$C_{10,\,\mu}^{\prime{\rm BSM}}$} & 
       \multirow{2}{*}{$0.01$} & \multirow{2}{*}{$0.10$} &  \multirow{2}{*}{$0.08$} 
        & $0.20$ & $0.19$  & $0.16$ & 
        \multirow{2}{*}{$0.1$} &  \multirow{2}{*}{$1.2$} & \multirow{2}{*}{$1.1$}   \\
  & & & & $-0.17$ & $0.01$ & $0.01$ & & & \\  \hline\hline
   \end{tabular}
\end{center}
\caption{\em As in table~\ref{tab:FARTFit} and table~\ref{tab:FARTFit2}, adding Moriond EW 2019  data about $R_K$ and $R_{K^*}$.
\label{tab:FARTFit2Update}}
\end{table}%

\footnotesize
 

\label{6out}

\end{document}